\documentclass[prd,nofootinbib,
, twocolumn
, nofootinbib
, superscriptaddress
]{revtex4-2}

\usepackage{aas_macros}
\usepackage{graphicx}
\usepackage[caption=false]{subfig}
\usepackage{siunitx}
\usepackage{mathrsfs}
\usepackage{amsmath,amssymb}
\usepackage{bm}
\usepackage{braket}
\usepackage{listings}
\usepackage{cases}
\usepackage{here}
\usepackage{comment}
\usepackage{soul}
\usepackage{cancel}
\usepackage{cases}
\usepackage[utf8]{inputenc}
\usepackage{url}
\usepackage{longtable}
\usepackage[normalem]{ulem}
\usepackage{xspace}
\usepackage[colorlinks=true
,urlcolor=blue
,anchorcolor=blue
,citecolor=blue
,filecolor=blue
,linkcolor=blue
,menucolor=blue
,linktocpage=true
,pdfproducer=medialab
,pdfa=true
]{hyperref}
\usepackage{animate}

\newcommand{\eq}[1]{\begin{equation}\begin{split} #1 \end{split}\end{equation}}

\newcommand{\Ms}{M_{\odot}}

\newcommand{\mr}{\mathrm}

\newcommand{\KA}[1]{\textcolor{magenta}{\sffamily [KA: #1]}}

\begin{document}

\title{CMB lensing from early-formed dark matter halos}
\author{Katsuya T. Abe}
\email{kabe@chiba-u.jp}
\affiliation{Center for Frontier Science, Chiba University, 1-33 Yayoi-cho, Inage-ku, Chiba 263-8522, Japan}
\author{Hiroyuki Tashiro}
\affiliation{Center for Education and Innovation, Sojo University, Ikeda, Nishi-ku, Kumamoto, 860-0082 Japan}

\begin{abstract}
Some theoretical models for the early Universe predict a spike-type enhancement in the primordial power spectrum on a small scale, which would result in forming early-formed dark matter halos~(EFHs).
Some recent studies have claimed to have placed limits on such small scales, which, however, involve uncertainties, such as the physics of substructures and the halo-galaxy relations.
In this work, we study the cosmic microwave background~(CMB) lensing effect, considering the existence of EFHs, and investigate the potential to probe the EFHs and the primordial perturbations on scales smaller than $1\mathrm{Mpc}$, complementing these previous studies.
We numerically calculate the angular power spectrum of the lensing potential and the lensed CMB anisotropy of temperature, E-mode, and B-mode polarization, including the nonlinear effects of EFHs. 
We find the possibility that the lensed CMB temperature anisotropy is significantly enhanced on small scales, $\ell>1000$, and could be tested by component decomposition of observed signals through multifrequency observations.
Through the calculation with different models of the spiky-type power spectrum, we demonstrate that the accurate measurements of the CMB lensing effect would provide insight into the abundance of EFHs within the limited mass range around $10^{11}~\Ms$ and the primordial power spectrum on the limited scales around $k\sim 1\mathrm{Mpc}^{-1}$.
In particular, we find that the existence of such EFHs can amplify the lensed anisotropy of CMB B-mode polarization even on large scales, $\ell <100$, as the overall enhancement by $\sim 5 \%$ level compared to the standard structure formation model without EFHs.
Therefore, future CMB measurements, such as the LiteBIRD satellite, can probe the existence of the EFHs and the spike-type primordial power spectrum through the precise measurement of the large-scale CMB B-mode polarization.
\end{abstract}

\maketitle

\section{Introduction}
The Universe exhibits rich hierarchical structures spanning vast ranges of magnitude, e.g., galaxy and galaxy clusters. The seeds of such hierarchical structures are called primordial perturbations. 
Precise measurements of cosmic microwave background~(CMB) and large-scale structures have been achieved to measure the amplitude of primordial perturbations on larger scales than $\mathcal{O}(0.1-1)$ Mpc~\cite{Planck2018_cospara,2006ApJS..163...80M,2019JCAP...07..017C,2021A&A...646A.129J,2021PhRvD.103d3522C,2021ApJS..253....3H,2021arXiv210513549D}.
They have also allowed us to reveal several of their statistical features on large scales, which remarkably align with predictions derived from the simplistic inflation scenario.
Conversely, the measurement of perturbations on smaller scales has encountered significant challenges, primarily due to the nonlinear effects associated with the formation of structures, although that is crucial to further understanding the inflationary mechanism.
Various scenarios, including inflation and dark matter, have proposed different types of small-scale fluctuation spectra.
For instance, several inflationary scenarios propose the occurrence of a spike-type enhancement, such as those found in single-field or multifield models featuring hybrid or double inflation~\cite{2015PhRvD..92b3524C, 2017PDU....18...47G, 2019JCAP...06..028B}.
Besides, particle production during inflation~\cite{PhysRevD.62.043508,PhysRevD.80.126018,PhysRevD.82.106009} and specific thermal histories such as an early matter-dominated era~\cite{2014JHEP...04..138B,PhysRevD.84.083503,2014PhRvD..90d3536F,PhysRevD.92.103505} or an era dominated by a fast-rolling scalar field~\cite{PhysRevD.98.063504} could generate a spike-type enhancement in the primordial power spectrum.

Such significantly enhanced small-scale fluctuations might play an essential role in forming cosmic structures and can largely impact cosmological signals. 
For example, the spike-type small-scale fluctuations predict the formation of primordial black holes and early-formed dark matter halos~(EFHs)\footnote{In previous works, EFHs with the spike-type excess have been discussed in the context of ultracompact minihalos~(UCMHs). However, UCMH originally indicates objects which gravitationally collapse near and by $z=1000$, so that requires $\delta \rho / \rho \gtrsim 10^{-3}$. On the contrary, we do not limit here the formation redshift of dark matter halos. Thus, we call them just EFHs.}.
Here, we investigate the scenario in which the power of fluctuations is still linear and focus on only EFHs.
This scenario has been examined and constrained through diverse cosmological signals. For instance, the energetic emission through DM annihilation from EFHs is one of the examples.
So far, the gamma-ray emission signal~\cite{2010PhRvD..82h3527J, 2009PhRvL.103u1301S,2012PhRvD..85l5027B,2016MNRAS.456.1394C,2018PhRvD..97b3539N,2018PhRvD..97d1303D} and the following effect on cosmic reionization~\cite{2011MNRAS.418.1850Z,2011JCAP...12..020Y,2016EPJP..131..432Y,2017JCAP...05..048C} have been investigated assuming that the dark matter is of a particle such as the weakly interacting massive particle~\citep{1985NuPhB.253..375S,1996PhR...267..195J,1999dmap.conf..592K}.
Nondetection of such energetic signals in Fermi-LAT~\cite{Atwood_2009} and Planck~\cite{2023JCAP...06..032A} and the observational value of the Thomson scattering optical depth
put constraints on the abundance of EFHs and the small-scale primordial scalar perturbation as $\mathcal{A}_{\zeta} < 10^{-7}$ for $10~{\rm Mpc}^{-1} <k < 10^8~{\rm Mpc}^{-1}$.
In addition, there are several works to examine the abundance in independent ways of the DM nature.
For example, Refs.~\cite{2012PhRvD..86d3519L,2016MNRAS.456.1394C} studied their gravitational lensing effect~\cite{2012PhRvD..86d3519L,2016MNRAS.456.1394C}, and Refs.~\cite{2020MNRAS.494.4334F,2022PhRvD.105f3531A,2022PhRvD.106h3521A} studied their baryonic effects. Both works have provided meaningful limits on the abundance of EFHs and also the small-scale primordial scalar perturbation, e.g., $\mathcal{A}_\zeta \lesssim 10^{-7}$
for $1~\mathrm{Mpc}^{-1} \lesssim k \lesssim 100~\mathrm{Mpc}^{-1}$ and ${\mathcal A}_{\zeta} \lesssim 10^{-6}$ on $100~{\rm
Mpc}^{-1}\lesssim k \lesssim 1000~{\rm Mpc}^{-1}$.

In this work, we focus on investigating the impact of these fluctuations on the weak gravitational lensing effect of CMB, commonly referred to as the CMB lensing effect. Understanding the influence of small-scale fluctuations on the CMB lensing effect is important to probe EFHs and primordial scalar perturbations on small scales in light of forthcoming observations of CMB anisotropies, particularly measurements of small-scale CMB temperature fluctuations, polarization modes, and large-scale polarization B-modes.

This paper is organized as follows.  Section~\ref{sec: pzeta_ucmh} describes
the additional power spectrum and the features of formed EFHs. 
In Sec.~\ref{sec: cmb_lensing_w_ucmh}, we discuss the CMB lensing effects considering the nonlinear effects of EFHs. In Sec.~\ref{sec: results}, we calculate the lensed anisotropies of the CMB temperature, E-mode, and B-mode polarization, including the nonlinear effects of EFHs and discuss the signals compared to the observation data. In Sec.~\ref{sec: conclude}, we conclude this paper.

Throughout this paper, we will assume a flat $\Lambda$-dominated cold dark matter~($\Lambda$CDM) cosmology and fix the cosmological parameters to the Planck 2018 best fits~\cite{Planck2018_cospara}, ($h=0.677$, $\Omega_{\mathrm{b}}h^2=0.0224, \omega_{\mathrm{cdm}}=0.120, \ln \left(10^{10} \mathcal{A}_{\zeta}\right)=3.05$, $n_{\mathrm{s}}=0.967$, and $\tau_{\text {reio }}=0.0544$)

\section{Early-formed dark matter halos from spike-type perturbations}\label{sec: pzeta_ucmh}

Our aim of this paper is to investigate 
the potential of the CMB gravitational lensing 
to probe the EFHs and their origin, \textit{i.e.,} primordial perturbations on scales smaller than 1~Mpc.
Therefore, in addition to the standard almost-scale-invariant
power spectrum,
we assume the presence of a spike-type power spectrum at a specific small scale in the primordial perturbation $\mathcal{P}_{\zeta}$,
\eq{
\label{eq:pk}
\mathcal{P}_{\zeta}
=\mathcal{P}^{\mr{st}}_{\zeta}(k)
+
\mathcal{P}^{\mr{add}}_{\zeta}(k).
}
Here $\mathcal{P}^{\mr{st}}_{\zeta}(k)$
denotes
the standard almost-scale-invariant
power spectrum,
$\mathcal{P}^{\mr{st}}_{\zeta}(k)
=A_\zeta (k/k_{\rm p})^{n_\mathrm{s}}$
with 
the amplitude
$\mathcal{A}_{\zeta} \sim 2.1\times 10^{-9}$,
the spectrum index $n_\mathrm{s}=0.967$,
and the pivot scale $k_{\rm p}= 0.05 \rm Mpc^{-1}$,
according to Ref.~\cite{Planck2018_cospara}.

In Eq.~\eqref{eq:pk}, $\mathcal{P}^{\mr{add}}_{\zeta}(k)$
is the additional spike-type power spectrum, which is given by
\eq{\label{eq: add_spike_pk}
\mathcal{P}^{\mr{add}}_{\zeta}(k)=
\frac{\mathcal{A}_{\zeta}^{\mathrm{add}}}{\sqrt{2\pi \sigma_{\mathrm{sp}}^2}} \exp \left[-\frac{1}{2}\left(\frac{\ln (k)-\ln \left(k_{\mathrm{sp}}\right)}{\sigma_{\mathrm{sp}}}\right)^2\right],
}
where $\mathcal{A}_{\zeta}^{\mathrm{add}}$ is an amplitude of the additional spike-type power spectrum, $k_{\mathrm{sp}}$ is a comoving wave number representing the peak scale of the spike, and $\sigma_{\mathrm{sp}}$ is a logarithmic width of the spike. In this work, we fix $\sigma_\mr{sp}=0.1$ for simplicity.
Thus, in the paper, we describe
the additional spike-type power spectrum
by using two parameters, $(k_{\mr{sp}},\mathcal{A}_{\zeta}^{\mathrm{add}})$.

The existence of the spike-type primordial perturbations provides an impact on the structure formation history of the Universe.
The spike-type primordial perturbation 
can induce a large amplitude in the matter density fluctuations and lead to the early formation of gravitational collapse objects, which are called EFHs. The EFHs produced by the spike-type power spectrum have been investigated in several works~\cite{2018PhRvD..97d1303D}

According to Refs.~\cite{2018PhRvD..97d1303D}, the scale density and radius of EFHs at their formed redshift $z_{\mathrm{coll}}$ are given by
\eq{
    r_s \simeq 0.7 \left[\left(1+z_{\mathrm{coll }}\right) k_{\mathrm{sp}}\right]^{-1},
}
\eq{
    \rho_s \simeq 30\left(1+z_{\mathrm{coll }}\right)^3 \bar{\rho}_{\mathrm{m,0}},
}
where $\bar{\rho}_{\mathrm{m, 0}}$ is the comoving mean matter density.
Thus, the enclosed mass at $z_{\mathrm{coll}}$ can be written by
\eq{\label{eq: ks_mass_relation_zcoll}
M_{\mathrm{coll}} & =4 \pi \rho_{\mathrm{s}} r_{\mathrm{s}}^3 m\left(u_\nu\right),
}
where $u_v \equiv r_{200} / r_{\mathrm{s}}$, $r_{200}$ is defined as the scale inside which the averaged dark matter density equals $200$ times the mean matter density of the Universe at $z_{\mathrm{coll}}$, and 
\eq{
m(x) & \equiv \int_0^x  \frac{u^{0.5}}{(1+u)^{1.5}} \mathrm{d} u.
}
We also incorporate the logarithmic mass increase due to gas accretion noted in Refs.~\cite{2018PhRvD..97d1303D}.
Finally, for a given $k_{\mathrm{sp}}$, we calculate the mass of EFHs as
\eq{\label{eq: ks_mass_relation}
M_{\mathrm{h}}(z; k_\mathrm{sp}) \approx 2\times 10^{9} \left(\frac{k_{\mathrm{sp}}}{10\mathrm{Mpc}^{-1}}\right)^{-3} \log\left(\frac{ 1+z_{\mathrm{coll}}}{1+z}\right)M_{\odot}.
}
Here, $z_{\mathrm{coll}}$ is calculated from $A_\zeta^{\mathrm{add}}$ and $k_\mathrm{sp}$. We will discuss the details later.

\begin{figure}[htbp]
    \centering
    \includegraphics[width=8cm,clip]{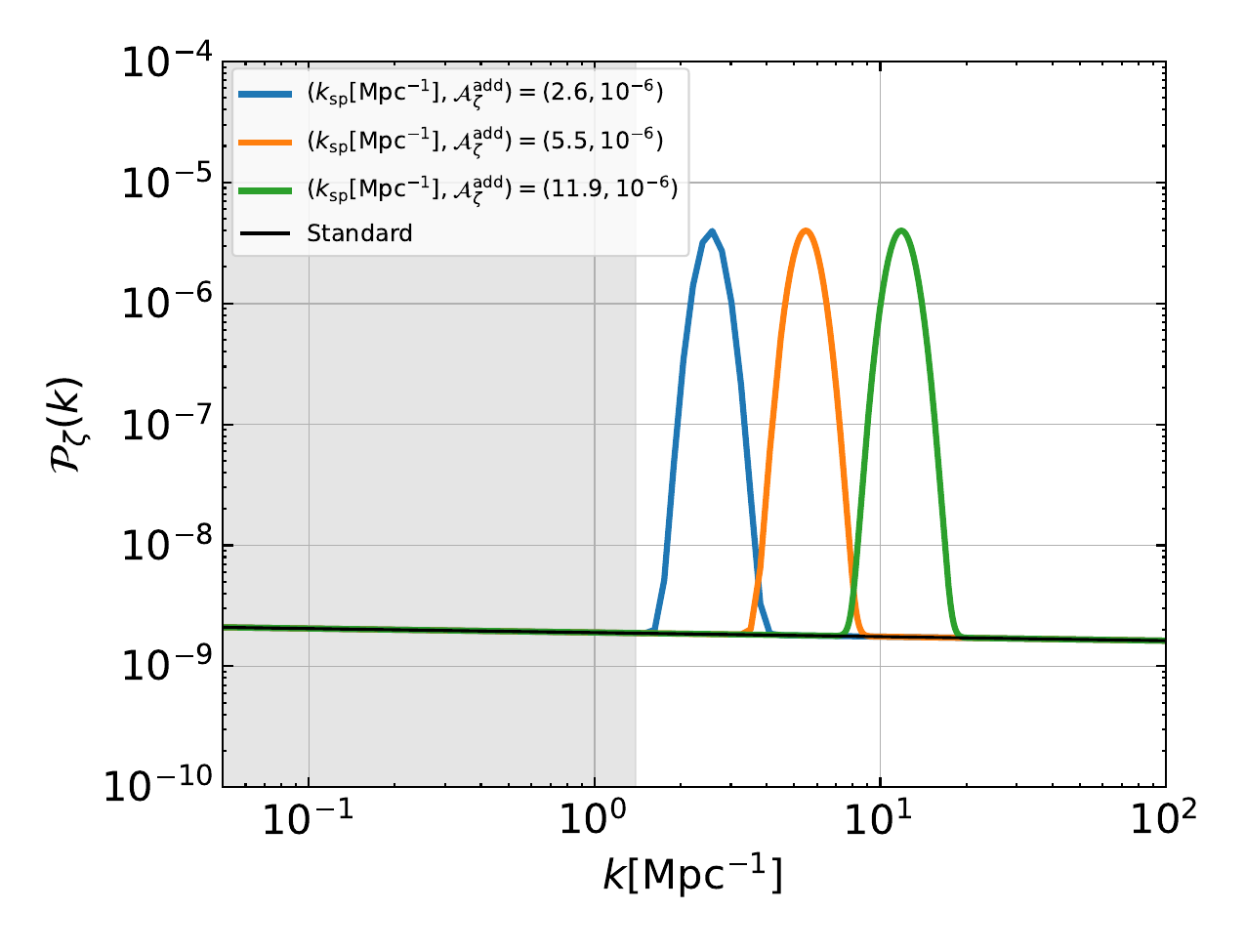}
    \caption{Three examples of the primordial power spectrum considered in this work. While a black solid line shows the standard almost-scale-invariant power spectrum with the amplitude of $\mathcal{A}_{\zeta} \sim 2.1\times 10^{-9}$, colored solid lines represent the ones with the additional spike-type power spectrum given by Eq.~\eqref{eq: add_spike_pk}. A location and amplitude of the spike are controlled by $k_{\mr{sp}}$ and $\mathcal{A}_{\zeta}^{\mathrm{add}}$. For these three examples, we fix as $\mathcal{A}_{\zeta}^{\mathrm{add}}=10^{-6}$ and set the three different spike scales with $k_{\mr{sp}}=2.6$~(blue), $5.5$~(orange), and $11.9~\mathrm{{Mpc}^{-1}}$~(green), respectively. The gray-shaded region indicates the scales at which other cosmological measurements can probe.}
    \label{fig: Pzeta_comparison}
\end{figure}

Figure~\ref{fig: Pzeta_comparison} shows primordial power spectra calculated from both the scale-invariant spectrum and three different spike-type spectra, which will be mainly examined as examples in this work.
For these spike-type power spectrums, we fix as $\mathcal{A}_{\zeta}^{\mathrm{add}}=10^{-6}$ and consider three models varying only k as examples;
$k_{\mr{sp}}=2.6$, $5.5$, and $11.9~\mathrm{{Mpc}^{-1}}$, which are shown in blue, orange, and green lines, respectively.
According to Eq.~\eqref{eq: ks_mass_relation}, these three values of $k_{\mr{sp}}$ correspond to the initial mass of the EFHs, $10^{11}$, $10^{10}$ and $10^{9}~M_\odot$, respectively.
We will discuss the dependency on $\mathcal{A}_{\zeta}^{\mathrm{add}}$ in Appendix~\ref{appendix: Azeta}.
The gray-shaded region in Fig.~\ref{fig: Pzeta_comparison} indicates the scales in which cosmological observation has already measured the primordial perturbations.
In the scales below the shaded region, 
there are only subtle constraints through measurements of CMB distortion by Far Infrared Absolute Spectrophotometer~(FIRAS) equipped with a Cosmic Background Explorer~(COBE)~\cite{1996ApJ...473..576F,1994ApJ...420..439M,2012ApJ...758...76C,2012MNRAS.425.1129C}, $\mathcal{P}_\zeta\lesssim 10^{-5}$, and nondetection of primordial black holes, $\mathcal{P}_\zeta\lesssim 10^{-2}$~\cite{2018JCAP...01..007E}. 
We should note that several more recent studies~\cite{2019MNRAS.489.2247C, 2022ApJ...928L..20S, 2022MNRAS.512.3163G, 2023arXiv230604674E} have claimed to have placed limits on matter density fluctuations on scales smaller than $1~\mathrm{Mpc}$. However, these studies involve some uncertainties, such as the physics of substructures and the halo-galaxy relations. Our study is not in conflict with these studies but complementary to the search for small-scale primordial fluctuations.

Now we discuss the number density of EFHs for a given spike-type power spectrum.
Since we consider here the EFHs formed from the spike-type fluctuations, the number density can be evaluated by employing the peak theory~\cite{1986ApJ...304...15B}~[hereafter Bardeen, Bond, Kaiser, and Szalay~(BBKS)] as
\eq{
\label{eq: num_dens_ucmh}
n(M_{\mr{h}},z)=\frac{k_{\mr{sp}}^{3}}{(2 \pi)^{2} 3^{3 / 2}} \int_{\delta_{c}/\mathcal{S}_{\mr{mat}}^{1 / 2}D_+(z)}^{\infty} e^{-\nu^{2} / 2} f(\nu) \mathrm{d} \nu,
}
where $\delta_{\mathrm{c}}=1.69$ is the linear density threshold for collapse, $\mathcal{S}_{\mr{mat}}$ is the present mass variance of the matter density fluctuation, $D_+(z)$ is the linear growth function at $z$, 
and $f(\nu)$ is the function which is defined as
\eq{
\begin{aligned}
f(\nu)= & \frac{1}{2} \nu\left(\nu^2-3\right)\left(\operatorname{erf}\left[\frac{1}{2} \sqrt{\frac{5}{2}} \nu\right]+\operatorname{erf}\left[\sqrt{\frac{5}{2}} \nu\right]\right) \\
& +\sqrt{\frac{2}{5 \pi}}\left\{\left(\frac{8}{5}+\frac{31}{4} \nu^2\right) \exp \left[-\frac{5}{8} \nu^2\right] \right.\\
& +\left. \left(-\frac{8}{5}+\frac{1}{2} \nu^2\right) \exp \left[-\frac{5}{2} \nu^2\right]\right\}.
\end{aligned}
}

The present mass variance $\mathcal{S}_{\mr{mat}}$ at the scale of $k_\mathrm{sp}$ is calculated from the power spectrum of the primordial perturbations in
\eq{
\label{eq:sigma_P}
\mathcal{S}_{\mr{mat}}(k_\mathrm{sp}) 
= \int d\log k~ \mathcal{P}_{\zeta }(k)T^2(k)  \tilde{W}_{k}^2(2\pi k /k_\mathrm{sp}),
}
where 
$\tilde{W}_k(x)$ represents the Fourier function of the window function, and $T(k)$ denotes the transfer function for the linear matter density fluctuations,
which can be 
obtained from {\tt CAMB}~\cite{Lewis:1999bs}.

Using this notation, the dimensionless linear matter density power spectrum at the present epoch is
described as
\eq{
\mathcal{P}^{\rm lin} _{\delta}(k) 
= \mathcal{P}_{\zeta }(k)T^2(k) 
\label{eq:mat_P}.
}
Note that, when calculating Eq.~\eqref{eq:sigma_P},
we make the approximation
$\mathcal{P}^{\mr{add}}_{\zeta}(k)=\mathcal{A}_{\zeta}^{\mathrm{add}}\delta(\ln(k)-\ln(k_{\mathrm{sp}}))$, for simplicity and employ the pointwise window function for $\tilde{W}_{k}$ following the BBKS.
Then the mass of the formed EFHs is limited to only $M_{\mr{h}}(k_{\mathrm{sp}})$ calculated in Eq.~\eqref{eq: ks_mass_relation}.
In addition, by differentiating Eq.~\eqref{eq: num_dens_ucmh} by the redshift, we can calculate the redshift at which EFHs are typically formed, $z_{\mathrm{typ}}\approx 2\mathcal{S}_{\mathrm{mat}}^{1 / 2}/\delta_{\mathrm{c}}-1$. For a given $k_\mathrm{sp}$ and $\mathcal{A}_\zeta^{\mathrm{add}}$, we fix $z_{\mathrm{coll}}$ represented in Eq.~\eqref{eq: ks_mass_relation} to $z_{\mathrm{typ}}$ in this work.

\section{Lensing potential with EFHs}\label{sec: cmb_lensing_w_ucmh}

To discuss the effect of EFHs on the CMB anisotropy, we focus on the CMB anisotropy power spectra for temperature~($C_\ell^{TT}$), E-mode~($C_\ell^{EE}$), and B-mode polarization~($C_\ell^{BB}$).
These lensed CMB observables
can be calculated through~\cite{2006PhR...429....1L}
\eq{\begin{aligned}\label{eq: lensed_cls}
& C_\ell^{\mr{L},TT}\hspace{-1mm} \approx\left(1-\ell^2 R^\phi\right) C_\ell^{TT}\\
&\ \hspace{0.8cm}+\int \frac{\mathrm{d}^2 \bm{\ell}^{\prime}}{(2 \pi)^2}\left[\bm{\ell}^{\prime} \cdot\left(\bm{\ell}-\bm{\ell}^{\prime}\right)\right]^2 C_{\left|\bm{\ell}-\bm{\ell}^{\prime}\right|}^{\phi\phi} C_{\ell^{\prime}}^{TT},
\\
& C_\ell^{\mr{L},EE}\hspace{-1mm}\approx\left(1-\ell^2 R^\phi\right) C_\ell^{EE}\\
& \hspace{0.8cm}+\int \frac{\mathrm{d}^2 \bm{\ell}^{\prime}}{(2 \pi)^2}\left[\bm{\ell}^{\prime} \cdot\left(\bm{\ell}-\bm{\ell}^{\prime}\right)\right]^2 C_{\left|\bm{\ell}-\bm{\ell}^{\prime}\right|}^{\phi\phi} C_{\ell^{\prime}}^{EE} \cos ^2 2\left(\phi_{\bm{\ell}^{\prime}}-\phi_{\bm{\ell}}\right), \\
& C_\ell^{\mr{L},BB}\hspace{-1mm}\approx\int \frac{\mathrm{d}^2 \bm{\ell}^{\prime}}{(2 \pi)^2}\left[\bm{\ell}^{\prime} \cdot\left(\bm{\ell}-\bm{\ell}^{\prime}\right)\right]^2 C_{\left|\bm{\ell}-\bm{\ell}^{\prime}\right|}^{\phi\phi} C_{\ell^{\prime}}^{EE} \sin ^2 2\left(\phi_{\bm{\ell}^{\prime}}-\phi_{\bm{\ell}}\right),
\end{aligned}
}
where $\bm{\ell}=(\ell, \phi_{\bm{\ell}})$, the superscript $\rm L$ denotes the 
lensed field and
$R^\phi$ is half of a total deflection angle power,
\eq{\label{eq: R_phi}
R^\phi \equiv \frac{1}{2}\left\langle|\nabla \phi|^2\right\rangle=\frac{1}{4 \pi} \int \frac{\mathrm{d} \ell}{\ell} \ell^4 C_\ell^{\phi\phi}.
}
Note that we neglect the primordial B-mode polarization. This would be justified by the current observational 
status so that the source of the primordial B-mode polarization is severely restricted,
e.g., in the term of the tensor-to-scalar ratio $r$,
$r<0.036$~\cite{PhysRevLett.127.151301}.

In Eq.~\eqref{eq: lensed_cls}, $C_\ell^{\phi\phi}$ is 
the angular power spectrum of the lensing potential, which is defined by

\eq{
\phi(\hat{\bm{n}}) \equiv-2 \int_0^{\chi_{\mr{S}}} \mathrm{~d} \chi \frac{\chi_{\mr{S}}-\chi}{\chi_{\mr{S}}\chi} \Phi\left(\chi \hat{\bm{n}} ; \eta_0-\chi\right),
}
where $\chi$ represents the comoving distance and $\chi_{\mr{S}}$ is the comoving distance to a source, \textit{i.e.,} the last scattering surface in this work.


Accordingly,
the angular power spectrum of the lensing potential 
can be calculated by~\cite{2006PhR...429....1L}
\eq{\label{eq: cellphiphi}
\begin{aligned}
C_\ell^{\phi\phi} &=16 \pi \int \frac{\mathrm{d} k}{k} \int_0^{\chi_{\mr{S}}} \mathrm{~d} \chi \int_0^{\chi_{\mr{S}}} \mathrm{~d} \chi^{\prime} \mathcal{P}_{\Phi}\left(k ; \eta_0-\chi, \eta_0-\chi^{\prime}\right) \\
& \hspace{1cm}\times j_\ell(k \chi) j_\ell\left(k \chi^{\prime}\right)
\left(\frac{\chi_{\mr{S}}-\chi}{\chi_{\mr{S}} \chi}\right)\left(\frac{\chi_{\mr{S}}-\chi^{\prime}}{\chi_{\mr{S}} \chi^{\prime}}\right)
\\
& \approx
 16 \pi \int \frac{\mathrm{d} k}{k} \int_0^{\chi_{\mr{S}}} \mathrm{~d} \chi 
 \mathcal{P}_{\Phi}\left(k ; \eta_0-\chi\right) 
\left(\frac{\chi_{\mr{S}}-\chi}{\chi_{\mr{S}} \chi}\right)^2,
\end{aligned}}
where $\mathcal{P}_{\Phi}\left(k ; \eta, \eta^{\prime}\right)$ is a dimensionless power spectrum of the gravitational potential $\Phi$ at $\eta$ and $\eta^\prime$ which is defined as
\eq{
\left\langle \tilde{\Phi}(\bm{k} ; \eta) \tilde{\Phi}^*\left(\bm{k}^{\prime} ; \eta^{\prime}\right)\right\rangle=\frac{2 \pi^2}{k^3} \mathcal{P}_{\Phi}\left(k ; \eta, \eta^{\prime}\right) \delta\left(\bm{k}-\bm{k}^{\prime}\right),
}
and $\mathcal{P}_{\Phi}\left(k ; \eta \right) \equiv
\mathcal{P}_{\Phi}\left(k ; \eta, \eta \right)$.
To obtain the last equation in Eq.~\eqref{eq: cellphiphi},
we apply the Limber approximation because 
in the calculation of the lensed CMB anisotropy,
the main contribution comes from the angular power spectrum of the lensing potential on small scales~$\ell \gtrsim 100$.

Using the Poisson equation, $\mathcal{P}_{\Phi}$ is related to a dimensionless matter power spectrum $\mathcal{P}_{\delta}$ as
\eq{\label{eq: poisson_pk}
\mathcal{P}_{\Phi}(k ; \eta)=\frac{9 }{4}\Omega_m^2(\eta) \mathcal{H}^4(\eta) 
\frac{\mathcal{P}_{\delta}(k ; \eta)}{k^4},
}
where $\mathcal{H}(\eta)$ is the conformal Hubble parameter with respect to $\eta$. One can find that the angular power spectrum of the lensing potential depends on the matter power spectrum $\mathcal{P}_{\delta}(k; \eta)$ through Eqs.~\eqref{eq: cellphiphi} and \eqref{eq: poisson_pk}.

In the calculation of the CMB lensing, the effect of the nonlinear structure, like DM halos, is not negligible. 
Therefore, it is required
to obtain the nonlinear matter power spectrum instead of the linear one 
as shown in Eq.~\eqref{eq:mat_P}
for a calculation of Eq.~\eqref{eq: poisson_pk}.
Similarly to the primordial perturbations as in Eq.~\eqref{eq:pk}, we can classify the nonlinear matter power spectrum into 
two parts as
\eq{
\label{eq:mpk}
\mathcal{P}_{\delta}
=\mathcal{P}^{\mr{st}}_{\delta}(k; \eta)
+
\mathcal{P}^{\mr{add}}_{\delta}(k; \eta).
}
Here, while $\mathcal{P}^{\mr{st}}_{\delta}(k; \eta)$ is 
the contribution to the matter power spectrum from the 
the standard almost-scale-invariant power spectrum,
$\mathcal{P}^{\mr{add}}_{\delta}(k; \eta)$ is the one from the 
spike-type power spectrum.

Though many references have studied such nonlinear effects 
due to the almost-scale-invariant power spectrum
accurately, we employ 
the {\tt HMC\small{ODE}} presented in Ref.~\cite{2016MNRAS.459.1468M} for the calculation of $\mathcal{P}^{\mr{st}}_{\delta}(k; \eta)$, including the nonlinear effects.

To obtain $\mathcal{P}^{\mr{add}}_{\delta}(k; \eta)$, 
it is necessary to consider the spike-type primordial power spectrum and the nonlinear effects induced by EFHs.
Following Ref.~\cite{2000MNRAS.318..203S}, 
the dimensionless matter power spectrum with the nonlinear effect can be written by 
\eq{
\label{def: matter_power}
\mathcal{P}^{\mr{add}}_{\delta}(k ; \eta)
=\mathcal{P}_{\delta}^{\mr{h h}}(k ; \eta)+\mathcal{P}_{\delta}^{\mr{P}}(k ; \eta),
}
where $P_{\delta}^{\mr{P}}$ is the Poisson term, and $P_{\delta}^{\mr{h h}}$ is the halo-halo correlation term.

In the case of EFHs, we obtain both terms using Eq.~\eqref{eq: num_dens_ucmh} as
\eq{\label{eq: 1h_term}
\mathcal{P}_{\delta}^{\mr{P}}(k;\eta)=\frac{k^3}{2 \pi^2}\frac{n(\eta)M_{\rm{h}}^2}{\bar{\rho}_{m}^2}|\tilde{y}[k, M_{\rm h}]|^2,
}
and
\eq{\label{eq: 2h_term}
\mathcal{P}_{\delta}^{\mr{h h}}(k;\eta)=D_+^2(\eta)\mathcal{P}^{\mr{lin}}_\delta (k)\left[\frac{n(\eta)M_{\rm{h}}}{\bar{\rho}_{m}} b_{\rm h} ~\tilde{y}[k, M_{\rm h}]\right]^2,
}
where $\mathcal{P}^{\mr{lin}}_\delta(k)$ is the present linear matter power spectrum given in Eq.~\eqref{eq:mat_P}, 
$b_{\mr{h}}$ is a bias factor for EFHs based on the peak theory~\cite{1997MNRAS.284..189M}, and
$\tilde{y}$ is a Fourier function of the density profile inside EFHs normalized by $M_{\mr{h}}$.

We should mention that we here neglect the evolution of the halo mass function due to the merger after the formation of EFHs, \textit{i.e.,} the sub-Poissonian clustering effect. It is possible that we overestimate their nonlinear effects, which we will leave for the next research.

For the density profile of EFHs,
we adopt the Navarro-Frenk-White~(NFW) profile,
\eq{
\rho(r)=\frac{\rho_\mathrm{s}}{\left(r / r_\mathrm{s}\right)\left(1+r / r_\mathrm{s}\right)^{2}},
}
where $\rho_{\mathrm{s}}$ is the characteristic density and $r_{\mathrm{s}}$ is the scale radius.
The scale radius relates to the concentration parameter, $c=r_{\mathrm{vir}}/r_s$ with a virial radius $r_{\mathrm{vir}}$. This concentration parameter generally depends on the mass and redshift, often referred to as the mass-concentration parameter relation. 
In this work, we adopt a fitting formula presented by Ref.~\cite{2014ApJ...789....1D} and updated parameters of mass-concentration relation~\cite{2019ApJ...871..168D,2020arXiv200714720I}. 

It is worth mentioning that there are some arguments and suggestions about the 
several density profiles for the EFHs, which are different from the NFW profile, based on the radial infall similarity solution~\cite{1985ApJS...58...39B} and numerical simulations~\cite{1999MNRAS.310.1147M, 2018PhRvD..97d1303D}.
However, since the density profile dependence arises below the scales of the EFH radius in the CMB lensing anisotropy,
the choice of the density profile does not affect our results
in our interesting scales, $\ell\lesssim \mathcal{O}(10^4)$.

\begin{figure}[htbp]
    \centering
    \includegraphics[width=8cm,clip]{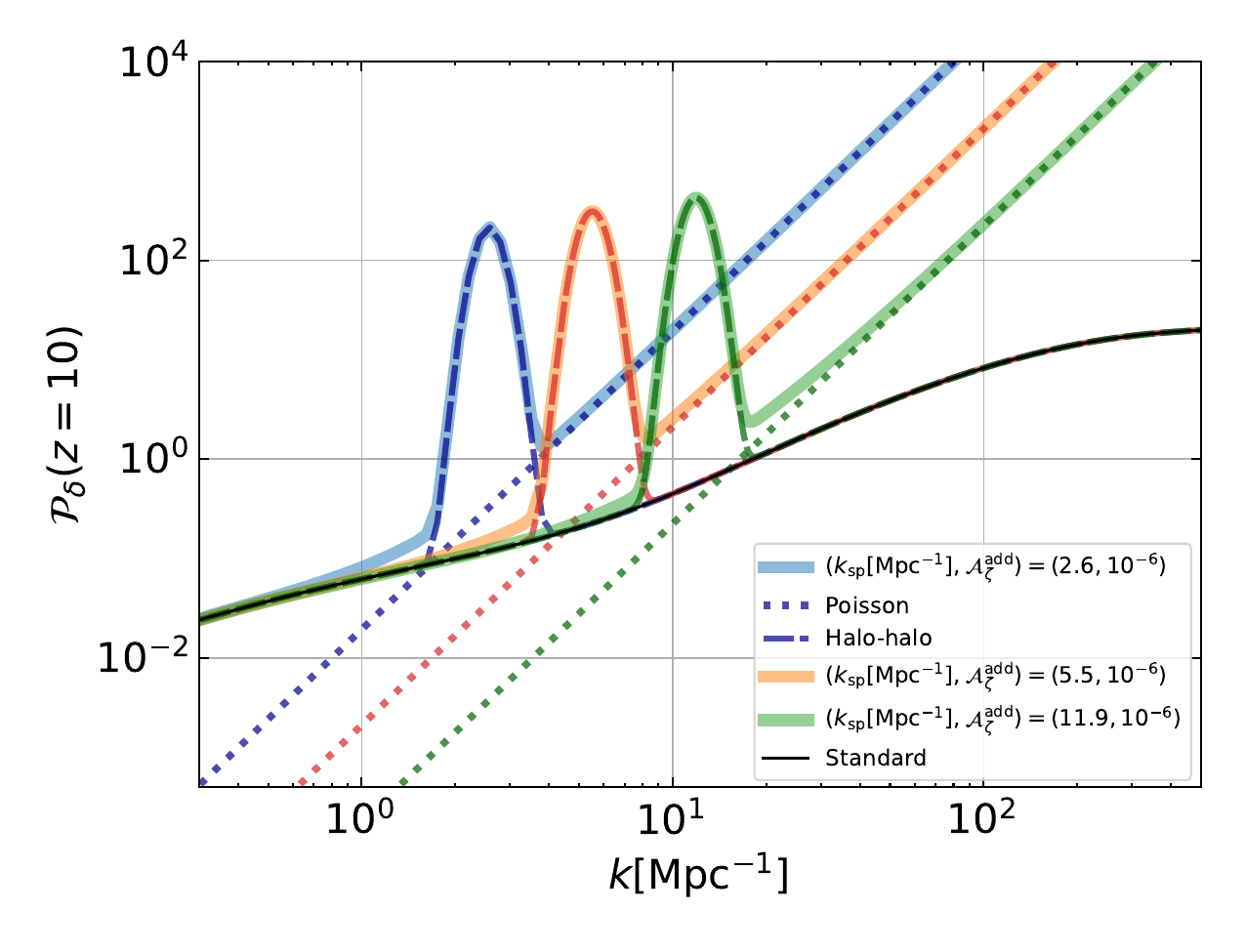}
    \caption{Matter power spectrum including the additional spike-type power and nonlinear effects of EFHs at $z=10$.
    We here fix as $\mathcal{A}_{\zeta}^{\mathrm{add}}=10^{-6}$ and plot the three different spectra with $k_{\mr{sp}}=2.6$, $5.5$, and $11.9~\mathrm{{Mpc}^{-1}}$ in blue, orange, and green solid lines, respectively. The dotted and dashed lines represent the Poisson term and the halo-halo correlation term of EFHs, respectively. For comparison, we also represent the nonlinear standard matter power spectrum,~$\mathcal{P}^{\mr{st}}_{\delta}(k; \eta)$, in the black solid line.}
    \label{fig: mPk_comparison}
\end{figure}

We calculate and plot the matter power spectrum with the additional spike-type power spectrum at $z=10$ in Fig.~\ref{fig: mPk_comparison}.
In the calculation, we include the nonlinear effects of EFHs, as discussed above.
The same manner of color as Fig.~\ref{fig: Pzeta_comparison} is adopted in Fig.~\ref{fig: mPk_comparison}.
The dotted and dashed lines
represent the Poisson term and the halo-halo correlation term of EFHs, respectively.
For comparison, we represent the nonlinear standard matter power spectrum,~$\mathcal{P}^{\mr{st}}_{\delta}(k; \eta)$, in the black solid line.
We find that the matter power spectrum is affected not only by the primordial spike-type power spectrum but also by the Poisson term as their nonlinear effect.

EFHs form earlier than the standard hierarchical structure 
generated from the scale-invariant spectrum $\mathcal{P}^{\mr{st}}_{\zeta}(k)$.
However as the Universe evolves, $\mathcal{P}^{\mr{st}}_{\zeta}(k)$ 
can produce more massive DM halos than EFHs.
In the formation of such massive halos, most of the EFHs would be captured into them.
After being captured, EFHs become subhalos or are destroyed by tidal disruption, and so on.
As a result, $\mathcal{P}^{\mr{add}}_\delta $ describing the nonlinear effect due to EFHs can be subdominant compared with $\mathcal{P}^{\mr{st}}_\delta $
in the late Universe.
To take the effect into account,
we introduce $z_{\mr{suv}}$ as the typical redshift
in which EFHs are captured into massive halos in the standard hierarchical structure formation.
We assume that, 
since EFHs cannot survive as isolated objects after $z_{\mr{suv}}$,
we neglect the nonlinear effects of EFHs in $\mathcal{P}_{\delta}$
after $z_{\mr{suv}}$.
We will discuss the value of $z_{\mr{suv}}$ later.

\begin{figure}[htbp]
    \centering
    \includegraphics[width=8cm,clip]{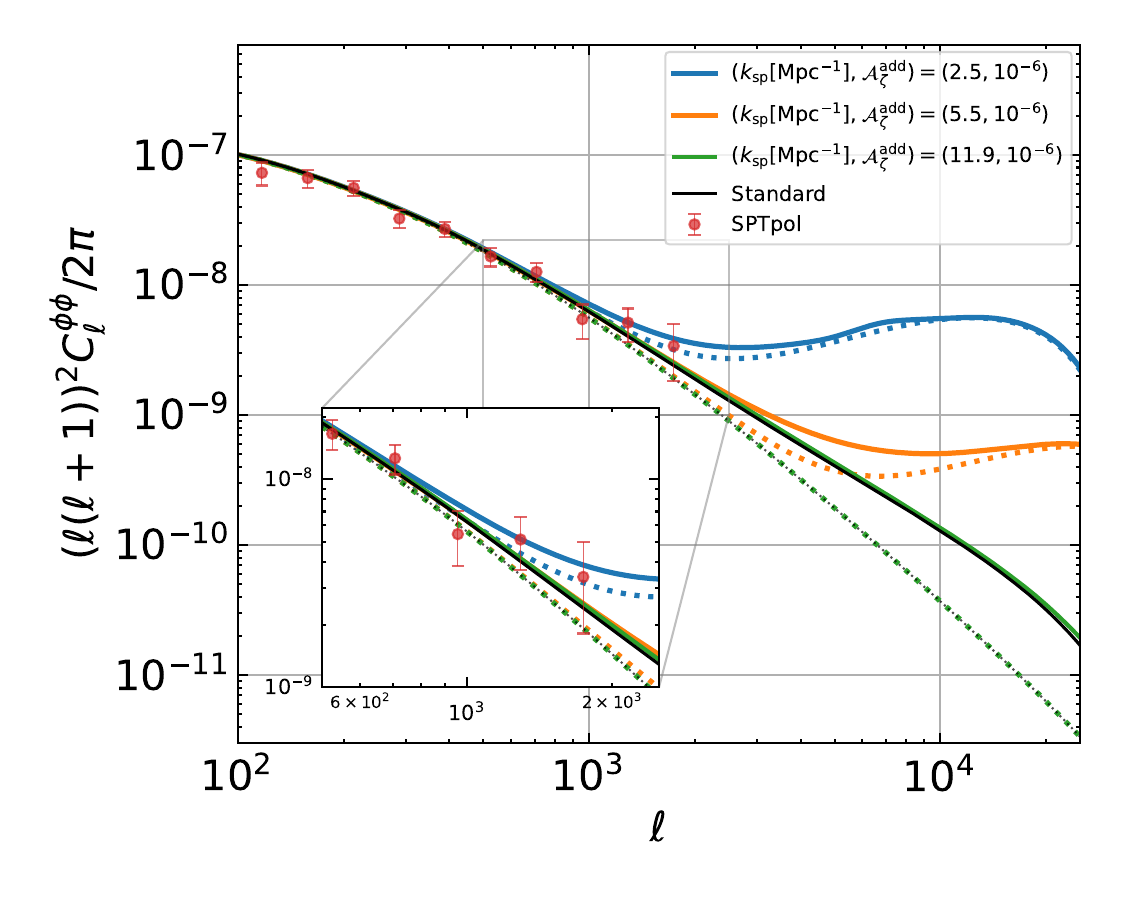}
    \caption{The angular power spectrum of lensing potentials including EFHs. The black solid line shows the standard one, including nonlinear effects from standard DM halos, while the black dotted line depicts the standard linear spectrum.
    The solid color lines represent the angular power spectrum, including the nonlinear effect with EFHs and standard DM halos. The color dotted lines depict the linear spectrum, only including the additional primordial power spectrum in Eq.~\eqref{eq: add_spike_pk}. The same manner of color in Figs.~\ref{fig: Pzeta_comparison} and \ref{fig: mPk_comparison} is adopted. For reference, we also plot the observation data by SPTpol~\cite{2020PhRvD.101l2003S} (red) with $\pm 1\sigma$ error bar.}
    \label{fig: clpp_comparison}
\end{figure}

Figure~\ref{fig: clpp_comparison} shows the angular power spectrum of lensing potential.
The black solid line shows the standard one, including nonlinear effects from standard DM halos. 
In contrast, the black dotted line depicts the standard linear spectrum.
The color solid lines represent the angular power spectrum, including the nonlinear effect with EFHs and standard DM halos, while the color dotted lines depict the linear spectrum, only including the additional primordial power spectrum in Eq.~\eqref{eq: add_spike_pk}.
The same manner of color in Figs.~\ref{fig: Pzeta_comparison} and \ref{fig: mPk_comparison} is adopted.
Here, when calculating the solid lines with $k_{\mr{sp}}=2.6$, $5.5$, and $11.9~\mathrm{{Mpc}^{-1}}$, we fix the value of $z_{\mr{suv}}$ based on the redshift at which the standard hierarchical structure formation, possessing a mass of $M=10^{11}~\Ms$ $M=10^{10}~\Ms$, and $M=10^{9}~\Ms$, becomes efficient, respectively; we set $z_{\mr{suv}}$ as $\delta_c/\sigma(M,z_{\mr{suv}})=1$, where $\sigma$ is the mass variance calculated from the standard matter power spectrum.
For instance,  we adopt $z_{\mr{suv}} \approx 3$ for $M=10^{11}~\Ms$~(blue), $z_{\mr{suv}}\approx 4$ for $M=10^{10}~\Ms$~(orange), and $z_{\mr{suv}}\approx 5$ for $M=10^{9}~\Ms$~(green), respectively.
We also plot the observation data by SPTpol~\cite{2020PhRvD.101l2003S} in red for comparison.

In Fig.~\ref{fig: clpp_comparison}, we find that the nonlinear effects give non-negligible enhancement in addition to the large bump induced by the spike-type enhancement in Eq.~\eqref{eq: add_spike_pk}.
However, we confirm that the main contribution to the nonlinear effect for these three models comes from standard halos rather than EFHs through additional calculations varying $z_\mathrm{suv}$, which is represented in Appendix~\ref{appendix: zsuv}.

We also put an enlarged figure of the $C_\ell^{\phi\phi}$s around $\ell=1000$.
Since the spectrum of $C_\ell^{\phi\phi}$ with different models matches on large scales, $\ell<1000$, their magnitude relationship around $\ell=1000$ decides the one of the total deflection angle power in Eq.~\eqref{eq: R_phi}.

\section{Results \& Discussion}\label{sec: results}

In order to calculate the lensed CMB anisotropy with the EFH effects, we employ the public Boltzmann code,
{\tt CAMB}~\cite{Lewis:1999bs}.
We modified {\tt CAMB} to consider the nonlinear effects by the existence of EFHs expressed in Eqs.~\eqref{eq: 1h_term} and~\eqref{eq: 2h_term}. 
Also, as we mentioned before, we employ analytical formulae of {\tt HMC\small{ODE}}~\cite{2016MNRAS.459.1468M} to calculate the nonlinear matter power spectrum with the standard structure formation scenario.

\begin{figure*}[t]
  \centering
  \includegraphics[width=0.3\linewidth]{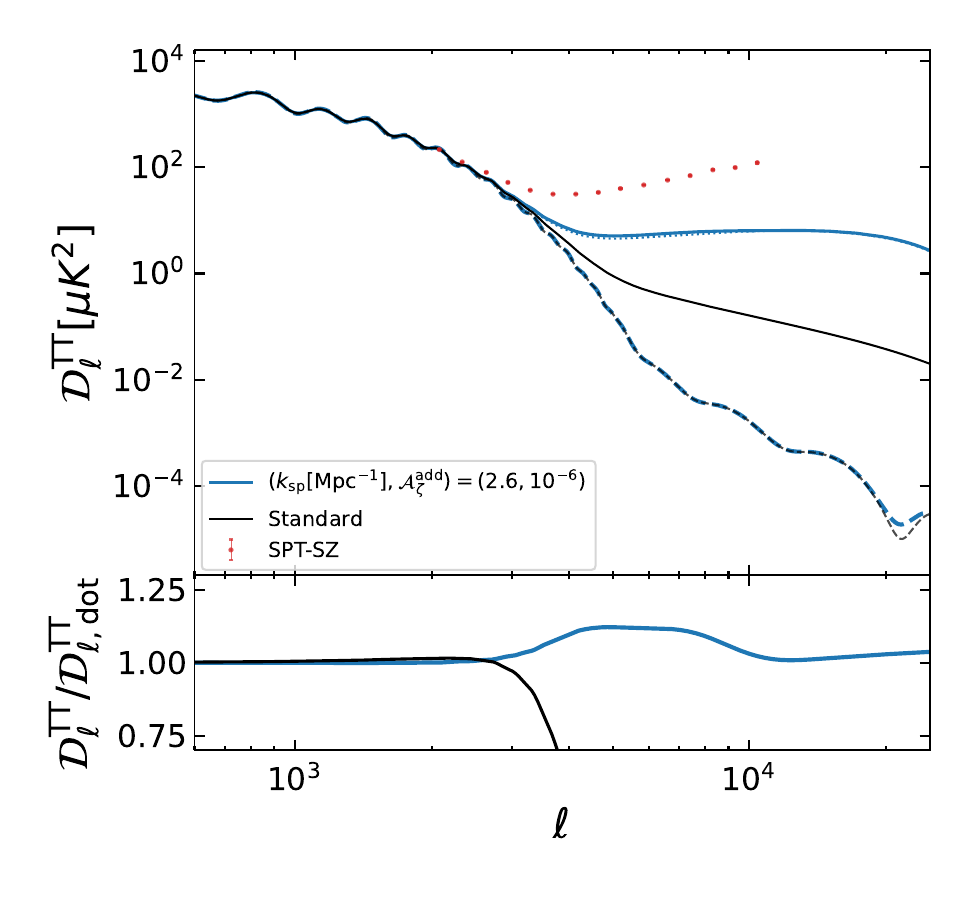}
  \hfill
  \includegraphics[width=0.3\linewidth]{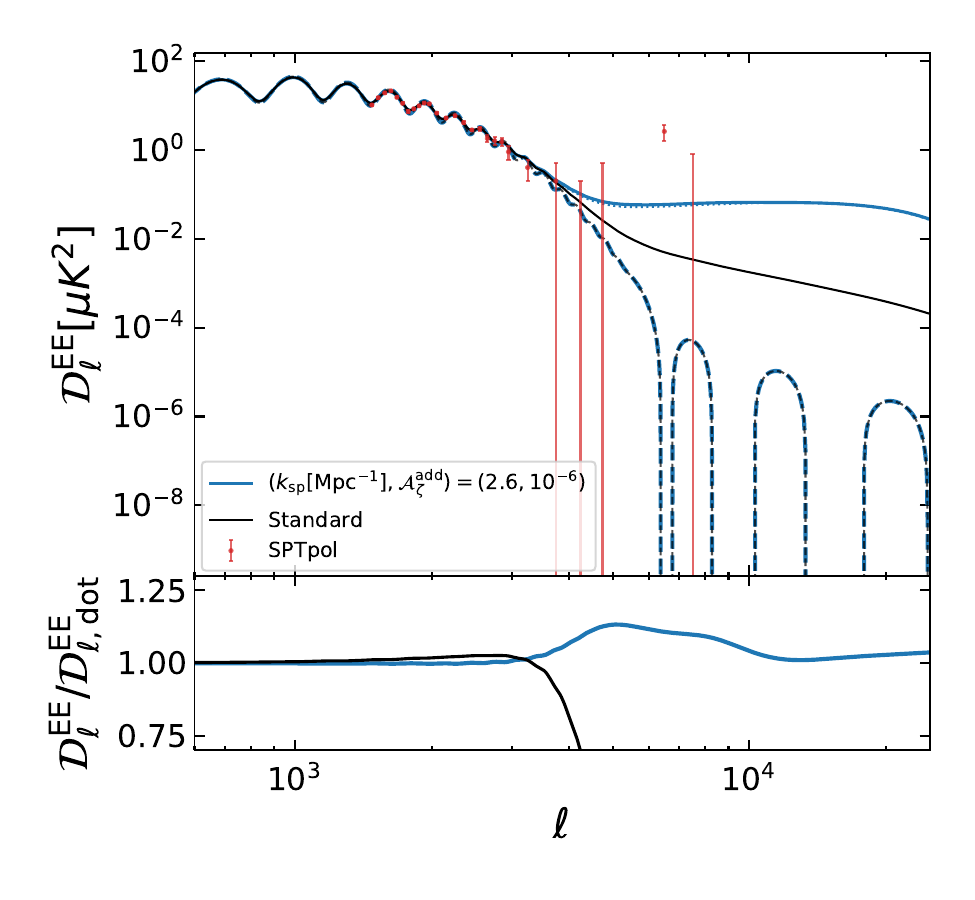}
  \hfill
  \includegraphics[width=0.3\linewidth]{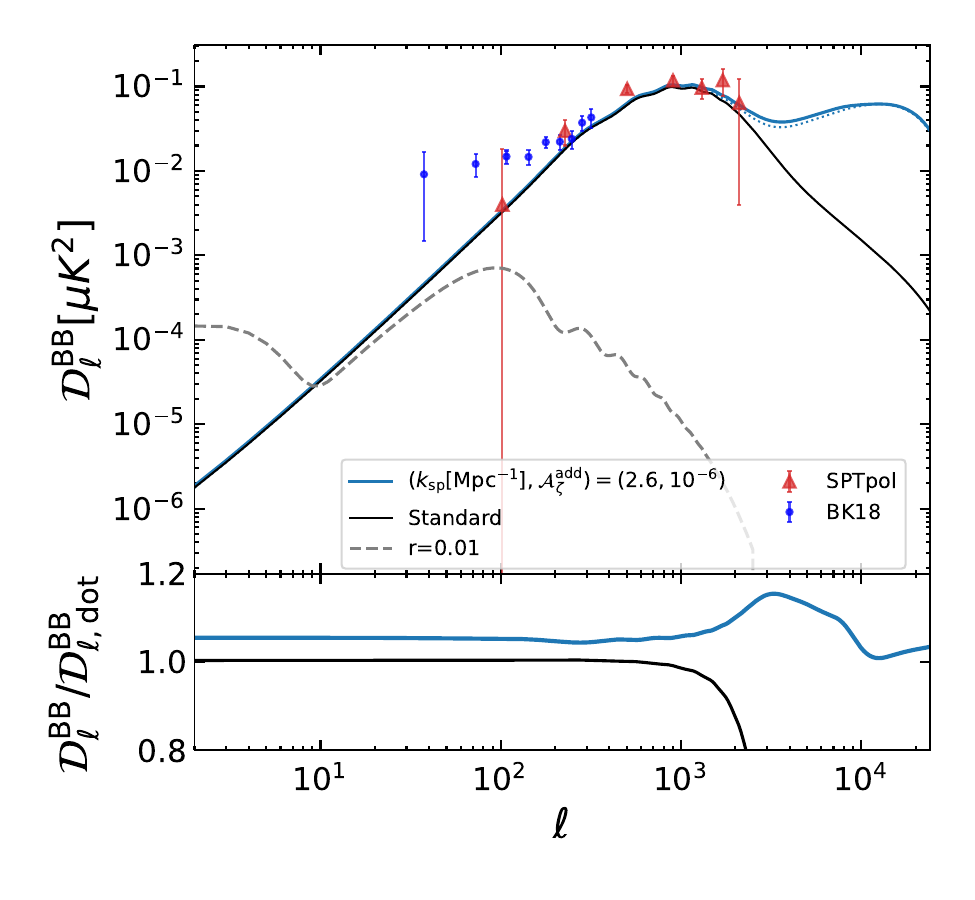}
  \caption{The lensed CMB angular power spectrum with the model of $(k_{\mr{sp}}~[\mr{Mpc}^{-1}],\mathcal{A}_{\zeta}^{\mathrm{add}}) =(2.6 , 10^{-6})$. The top of each panel shows the lensed angular power spectrum with respect to $D_\ell$ in the unit of $(\mu \mathrm{K})^2$. The black lines show the spectrum with standard DM halos, and the blue lines depict the ones with this model. The solid line shows the spectrum, including the nonlinear effect, while the dotted line represents the linear spectrum. The dashed lines represent the unlensed spectrum. The bottom of each panel represents the ratio of the blue and black solid lines against the blue dotted line. For reference, we also plot the observation data by SPT-SZ~\cite{2021ApJ...908..199R}, SPTpol ~\cite{2018ApJ...852...97H,2020PhRvD.101l2003S}  and BICEP/KECK~\cite{PhysRevLett.127.151301} with $\pm 1\sigma$ error bars. \textit{Left}: CMB temperature, \textit{Middle}: CMB E-mode polarization, \textit{Right}: CMB B-mode polarization.}
  \label{fig: cls_M1e12}
\end{figure*}

The left panel of Fig.~\ref{fig: cls_M1e12} shows the results of the lensed CMB temperature anisotropy with the EFH effect.
In the upper part, the vertical axis shows $\mathcal{D}_\ell^{TT}$ in the unit of $\mu \mathrm{K}^2$, where $\mathcal{D}_\ell^{XX}\equiv \ell(\ell+1)C_\ell^{XX}$.
The blue lines show the anisotropies related to the model of $(k_{\mr{sp}}=2.6 ~\mr{Mpc}^{-1},\mathcal{A}_{\zeta}^{\mathrm{add}}=10^{-6})$, while the black lines show the standard anisotropies. 
The black solid line depicts the standard lensed anisotropy, while the blue solid line shows the lensed anisotropy, including the additional power spectrum and nonlinear effects from standard halos and EFHs.
For the comparison, we also plot the linear lensed anisotropy for this model in the blue dotted line and the unlensed anisotropy in the blue dashed line.
The black dashed line represents the standard unlensed anisotropy.
In addition, we plot the observation data by SPT-SZ~\cite{2021ApJ...908..199R} for reference.
In the lower part, we plot the ratio of the blue and black solid lines against the blue dotted line.

The lensing effect enhances the amplitude of the anisotropy on small scales. 
We find that the lensed anisotropy with the nonlinear effects of EFHs and standard DM halos, \textit{i.e.,} the blue solid line, would be comparable to the thermal SZ effect by galaxy clusters or other sources, including dusty, star-forming, and radio galaxies, although it does not reach the level of the observed signal.
These signals have different frequency dependencies, while our lensed CMB anisotropy is independent of the frequency.
Component decomposition through multifrequency observations would provide more stringent constraints on our lensed signal and, subsequently, on the abundance of the EFH and the feature of the spike-type power spectrum.


We also calculate the lensed anisotropy of the CMB polarization.
The middle panel of Fig.~\ref{fig: cls_M1e12} shows the lensed anisotropy of the CMB E-mode polarization $D_\ell^{\mr{L}, EE}$.
The plotting manner is the same as in the left panel of Fig.~\ref{fig: cls_M1e12}.
We also plot the observation data by SPTpol ~\cite{2018ApJ...852...97H} for comparison.
Similarly to the temperature anisotropy, 
the lensing effect of the EFHs arises on small scales as the enhancement of the anisotropy signals.
According to the current observation data of the E-mode polarization, it is difficult to provide the constraint on the abundance of the EFHs due to the large error bars of the E-mode polarization measurement.
However, the figure suggests that the improvement of the observation measurement could provide the limit on the EFH abundance and the spike-type power spectrum of $(k_{\mr{sp}}=2.6 \mr{Mpc}^{-1},\mathcal{A}_{\zeta}^{\mathrm{add}}=10^{-6})$.

The right panel of Fig.~\ref{fig: cls_M1e12} represents the lensed anisotropy of the CMB B-mode polarization $D_\ell^{\mr{L},BB}$ 
with the same manner 
as in the left and middle panels.
Additionally, we plot the observation data by SPTpol~\cite{2020PhRvD.101l2003S} and BICEP/KECK~\cite{PhysRevLett.127.151301} in red and blue points, respectively.
As mentioned before, we calculate the lensed anisotropies, assuming that the primordial (unlensed) anisotropy of B-mode polarization can be negligible.
However, we plot the primordial B-mode polarization anisotropy with the tensor-to-scalar ratio of $r=0.01$ in the black dotted line for comparison.

In the B-mode polarization, the enhancement due to the spike-type power spectrum arises on small scales, as in
the temperature and E-mode polarization.
Since there are no sources of B-mode polarization other than the gravitational lensing in our calculation, the enhancement leads to the significant feature of the B-mode polarization on small scales.

Besides, the lensed B-mode signals are calculated by integrating $C^{\phi\phi}_\ell$ overall range of $\ell$ modes.
Therefore, although the existence of the EFHs amplify 
$C^{\phi\phi}_\ell$ on small scales, this amplification affects all range of $\ell$ modes in the lensed B-mode through the integration.
As a result, the effect of the spike-type power spectrum on the B-mode polarization appears even on large scales 
as the overall enhancement by $5~\%$ level.

Note that degeneracy between the flat enhancement and the amplitude of the standard primordial power spectrum $\mathcal{A}_\zeta$ can break using, e.g., CMB temperature and E-mode polarization anisotropies on scales of $\ell\leq 10^3$. Actually in the Planck analysis, the value of $\mathcal{A}_\zeta$ is obtained in the $\sim 5\%$ accuracy, $\mathcal{A}_{\zeta}=2.1^{+0.1}_{-0.1}\times 10^{-9}$.
While current CMB polarization measurements on large scales are constrained by instrumental noise, forthcoming projects such as CLASS~\cite{2022ApJ...926...33D}, Groundbird~\cite{2021ApJ...915...88L}, and PIPER~\cite{PIPER:2016szc}, can address the B-mode polarization anisotropies accurately, with limitations primarily imposed by the cosmic variance.
Notably, the LiteBIRD satellite~\cite{2020SPIE11443E..2FH} planned to be launched in the late 2020s will conduct polarimetric observations across the entire sky, employing 15 frequency bands to address the foreground predicament.
Therefore, though the principal objective of these experiments is to detect the primordial B-mode anisotropy, our model can be testable through these future experiments.

\begin{figure*}[t]
  \centering
  \includegraphics[width=0.3\linewidth]{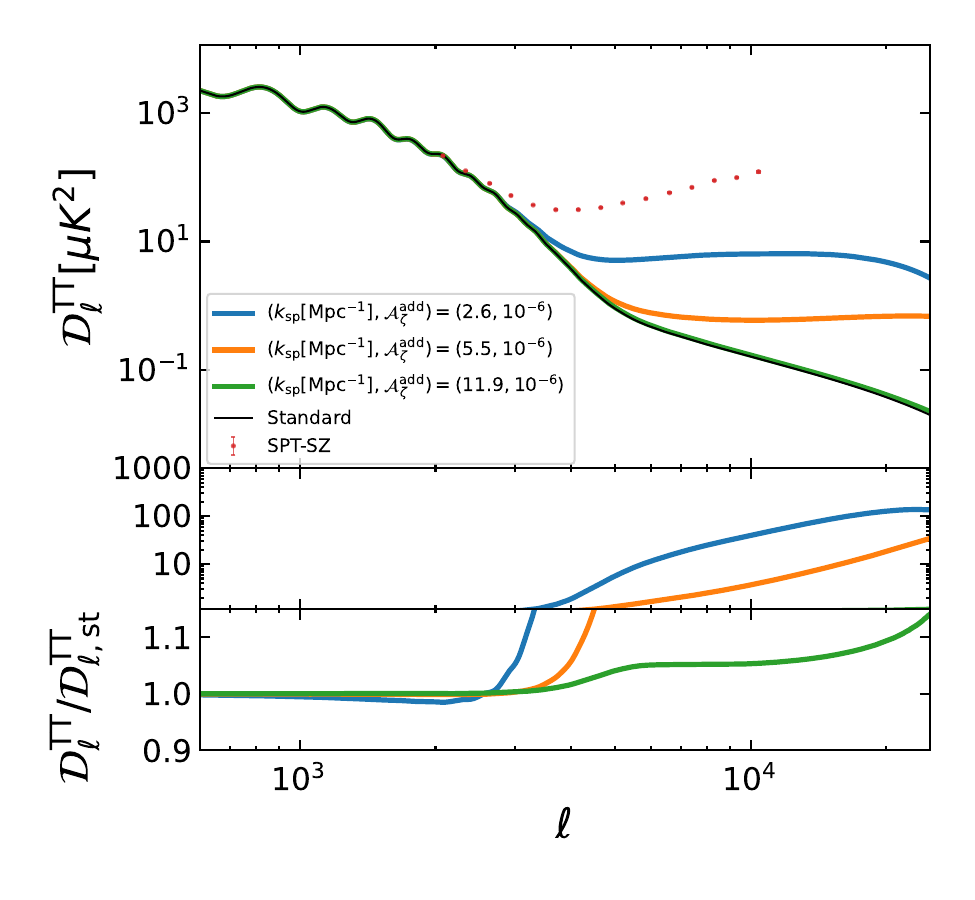}
  \hfill
  \includegraphics[width=0.3\linewidth]{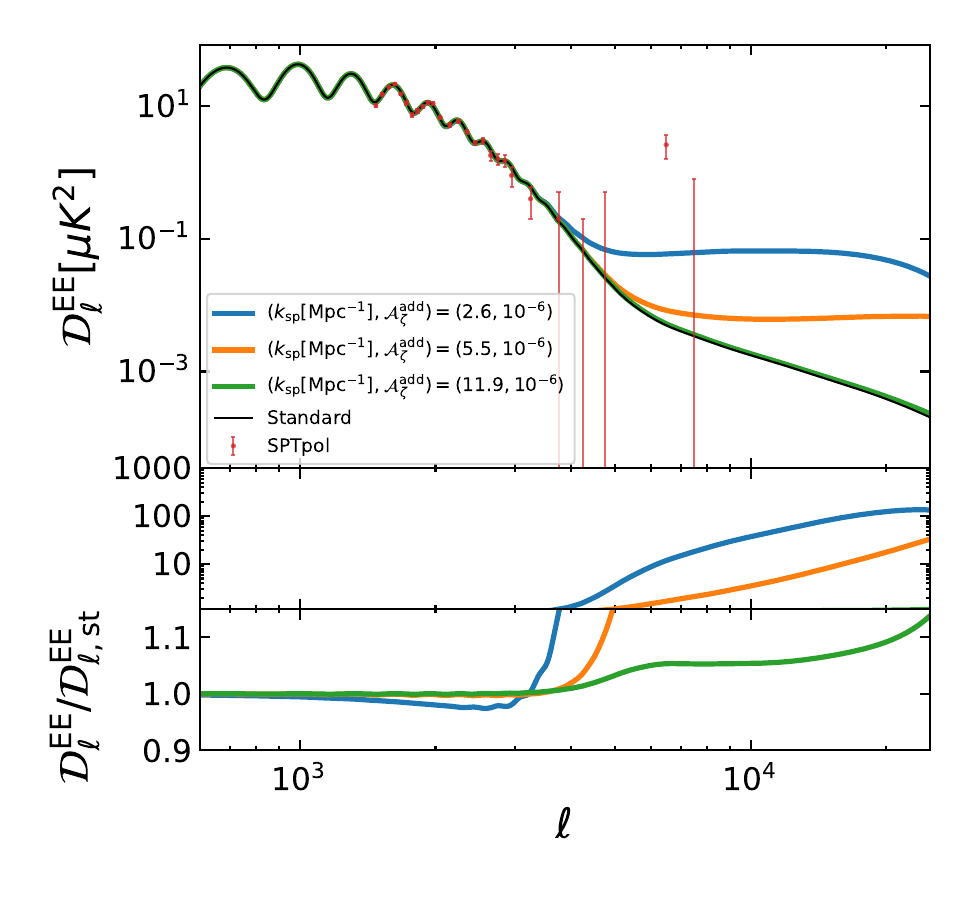}
  \hfill
  \includegraphics[width=0.3\linewidth]{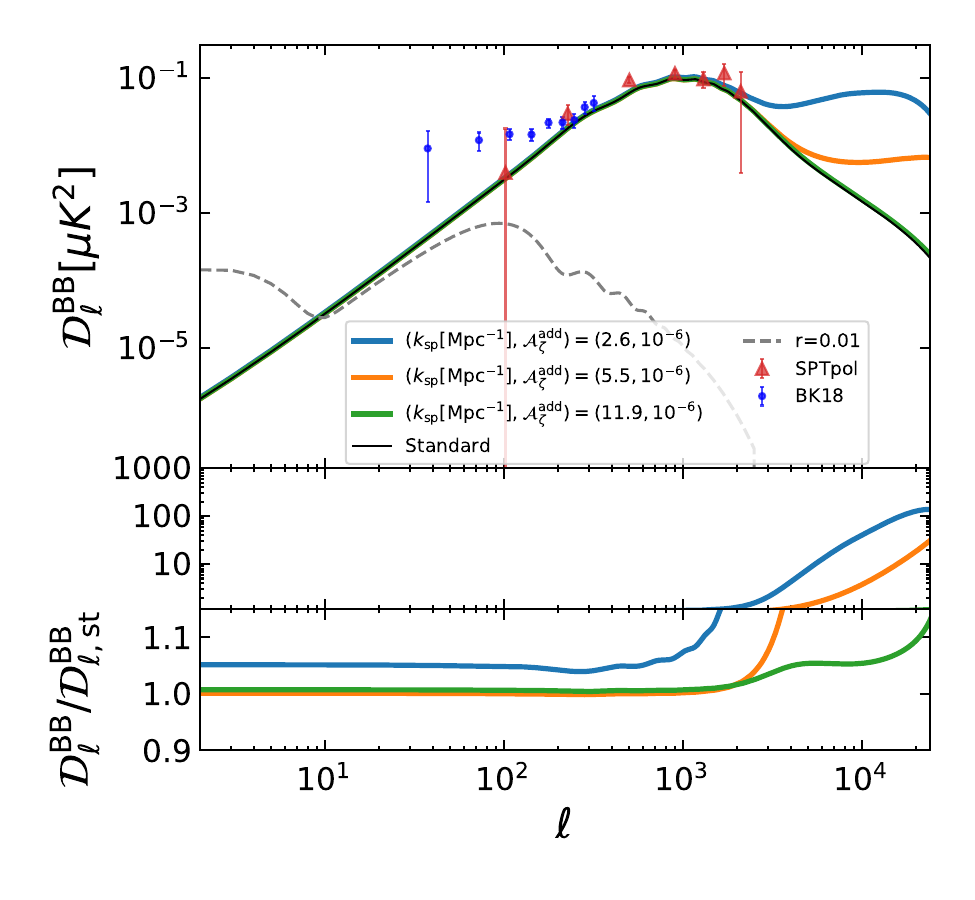}
  \caption{The lensed CMB angular power spectrum with the three models of $(k_{\mr{sp}}[\mr{Mpc}^{-1}],\mathcal{A}_{\zeta}^{\mathrm{add}}) = (2.6 , 10^{-6})$, $(5.5, 10^{-6})$, and $(11.9, 10^{-6})$. 
  At the top of each panel in Fig.~\ref{fig: cls_M1e10}, we plot the lensed CMB angular power spectrum, including the nonlinear effect with EFHs and standard DM halos as in Fig.~\ref{fig: cls_M1e12}.
  We plot these panels in the same manner of color as in Figs.~\ref{fig: Pzeta_comparison} and \ref{fig: mPk_comparison}.
  At the bottom of each panel, we show the ratio of the blue, orange, and green solid lines against the black solid line.  \textit{Left}: CMB temperature, \textit{Middle}: CMB E-mode polarization, \textit{Right}: CMB B-mode polarization.}
  \label{fig: cls_M1e10}
\end{figure*}

Lastly, we mention the results of adapting the other two models of $k_{\mr{sp}}=5.5$, and $11.9~\mathrm{{Mpc}^{-1}}$ to see the dependence of $C_\ell^{\mr{L}}$ on $k_{\mr{sp}}$, which is plotted in Fig.~\ref{fig: cls_M1e10}.
At the top of each panel in Fig.~\ref{fig: cls_M1e10}, we plot the lensed CMB angular power spectrum, including the nonlinear effect with EFHs and standard DM halos as in Fig.~\ref{fig: cls_M1e12}.
We plot these panels in the same manner of color as in Figs.~\ref{fig: Pzeta_comparison} and \ref{fig: mPk_comparison}.
At the bottom of each panel, we show the ratios of the solid lines against the black solid line.
As in the case of the parameter set of $(k_{\mr{sp}}[\mr{Mpc}^{-1}],\mathcal{A}_{\zeta}) = (5.5, 10^{-6})$, the nonlinear effects and the presence of a spike-type primordial power spectrum can make some impact on the lensed anisotropy. However, as expected from Fig.~\ref{fig: clpp_comparison}, the magnitude of the enhancement is too small to detect, mainly because of the peak location of the additional spike-type primordial power spectrum.
In the model of $(k_{\mr{sp}}~[\mr{Mpc}^{-1}],\mathcal{A}_{\zeta}) = (11.9, 10^{-6})$, the difference no longer appears in the lensed CMB anisotropy, as expected from Fig.~\ref{fig: clpp_comparison}.
Combining the dependency of $(\ell(\ell+1))^2 C_\ell^{\phi\phi}\propto \ell^{-2}$ and Eq.~\eqref{eq: lensed_cls}, one can find that the enhancement of $C_\ell^{\phi\phi}$ on small scales, $\ell>10^4$, no longer affects the lensed CMB anisotropies.
Actually, in cases of $k_{\mathrm{sp}}=5.5$ and $11.9~\mathrm{Mpc}^{-1}$, we confirmed that setting higher amplitude of $\mathcal{A}_\zeta^{\mathrm{add}}$ does not change the result.
Hence, we find that the accurate measurements of the anisotropy of lensing potentials and the lensed CMB anisotropy would provide insight into the abundance of EFHs with the mass of $\sim 10^{11}~\Ms$ and the primordial power spectrum on the limited scales around $k \sim 1 \mathrm{Mpc}^{-1}$.



\section{conclusion}\label{sec: conclude}
Some theoretical models for the early Universe, including the inflationary expansion scenario, predict a spike-type enhancement in the primordial power spectrum on small scales.
Such enhanced fluctuations form EFHs, which could significantly impact the structure formation and cosmological signals.
In this study, we have studied the CMB gravitational lensing effect, considering the existence of the spike-type primordial power spectrum and the nonlinear effects of EFHs, and investigated the potential to probe the EFHs and their origin, \textit{i.e.,} primordial perturbations on scales smaller than 1~Mpc.

This work has investigated three different spike-type power spectra with $k_{\mr{sp}}[\mr{Mpc}^{-1}]=2.6$, $5.5$, and $11.9$ while fixing $\mathcal{A}_{\zeta}^{\mathrm{add}} = 10^{-6}$ as examples.
We have first derived the matter power spectrum incorporating the nonlinear effects of EFHs, as shown in Fig.~\ref{fig: mPk_comparison}. We then calculated the angular power spectrum of the lensing potential, which is plotted in Fig.~\ref{fig: clpp_comparison}. We have found that although the nonlinear effect, especially the Poisson term of EFHs, appears on the matter power spectrum, it is less effective in the angular power spectrum of the lensing potential. We have also found that the EFHs contribution is smaller than that of standard halos in the three different spike-type power spectra we investigated.
After that, we numerically calculated the lensed CMB anisotropy of temperature, E-mode, and B-mode polarization for the two power spectra.

We have found that the effect of the spike-type power spectrum and EFHs on the CMB observables arises on small scales, as shown in Figs.~\ref{fig: clpp_comparison}, \ref{fig: cls_M1e12} and~\ref{fig: cls_M1e10}.
Regarding the lensed anisotropies of the CMB temperature and E-mode polarization with $(k_{\mr{sp}}[\mr{Mpc}^{-1}],\mathcal{A}_{\zeta}^{\mathrm{add}})=(2.6, 10^{-6})$, we have found the possibility that the signals are significantly enhanced and could be discussed by component decomposition of observed signals through multifrequency observations or improvements of the measurements.
As for the lensed anisotropy of CMB B-mode polarization, we have found that the effect of the spike-type power spectrum and EFHs appears even on large scales, $\ell <100$, as the overall enhancement by $\sim 5 \%$ level compared to the standard one. Although the current CMB observation does not have the sensitivity to distinguish $5\%$ difference on the scales, the future measurements on CMB B-mode polarization on large scales, such as the LiteBIRD satellite, have the potential to test the existence of EFHs and the spike-type power spectrum on small scales.

On the other hand, we have found that the enhancements by the effect of the spike-type power spectrum and EFHs with $(k_{\mr{sp}}[\mr{Mpc}^{-1}],\mathcal{A}_{\zeta}^{\mathrm{add}})=(5.5,10^{-6})$ and $(11.9,10^{-6})$ are too small to detect mainly because of the peak location of the additional spike-type primordial power spectrum.
Therefore, we have found that the accurate measurements of the CMB lensing effect would provide insight into the abundance of EFHs within the limited mass range around $10^{11}~\Ms$ and the primordial power spectrum on the limited scales around $k\sim 1\mathrm{Mpc}^{-1}$.

Finally, we leave some comments on the EFH exploration. As we have investigated, the nonlinear effects of EFHs are difficult to see in the CMB weak lensing effect, which may suggest that it is more important to search for individual gravitational lensing events by EHFs, as has been done in Ref.~\cite{2012PhRvD..86d3519L}, to search for EFHs. In addition, since the effect might still be strongly present in the matte power spectrum, it would be useful to use, e.g., the intensity mapping of the 21cm line to probe EFHs in the future.

\hspace{3mm}
\acknowledgments
We thank V. Nistane for the useful discussions and for the cooperation in developing the modified CAMB code.
This work was supported by JSPS KAKENHI Grant No. 21K03533.

\appendix
\section{$\mathcal{A}^{\mathrm{add}}_{\zeta}$ dependency}\label{appendix: Azeta}
\begin{figure*}[htbp]
    \begin{tabular}{cc}
      \begin{minipage}[t]{.5\linewidth}
        \centering
        \includegraphics[width=.9\hsize]{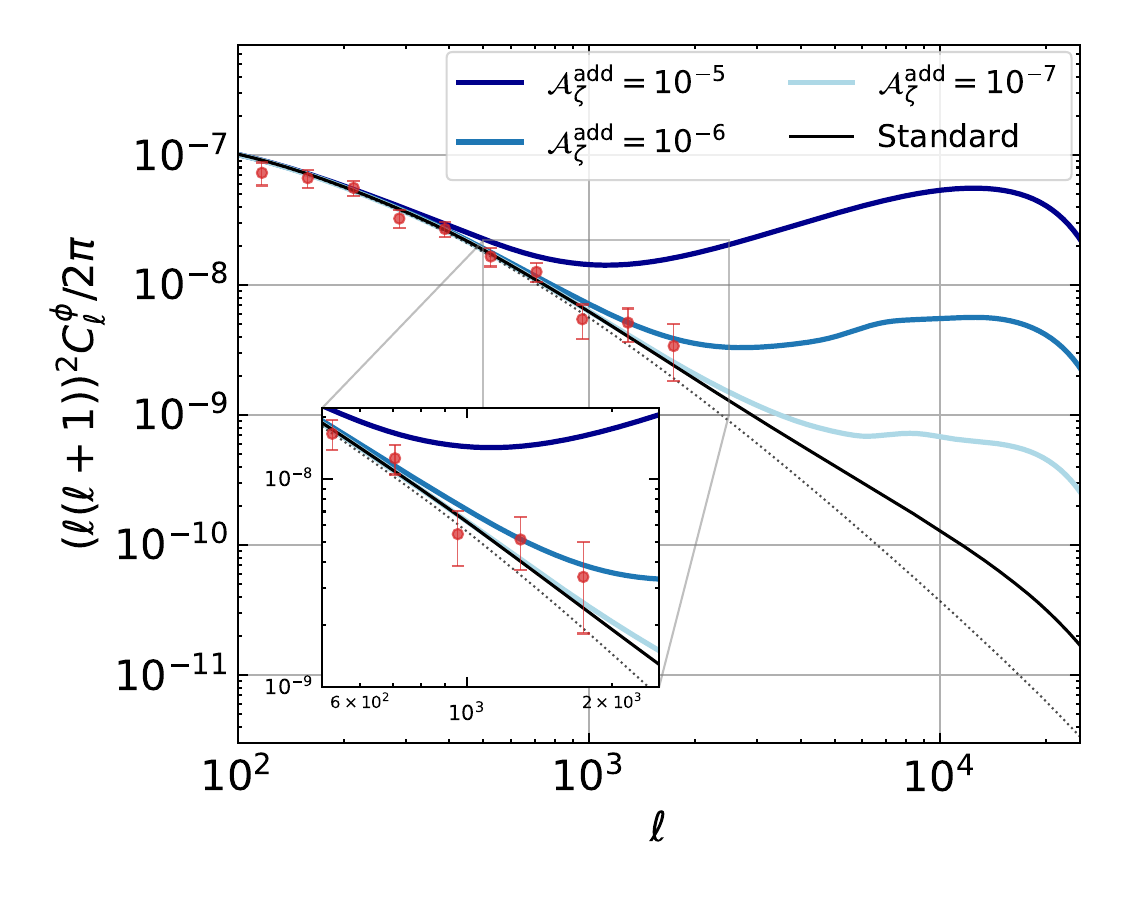}
      \end{minipage} &
      \begin{minipage}[t]{.5\linewidth}
        \centering
        \includegraphics[width=.9\hsize]{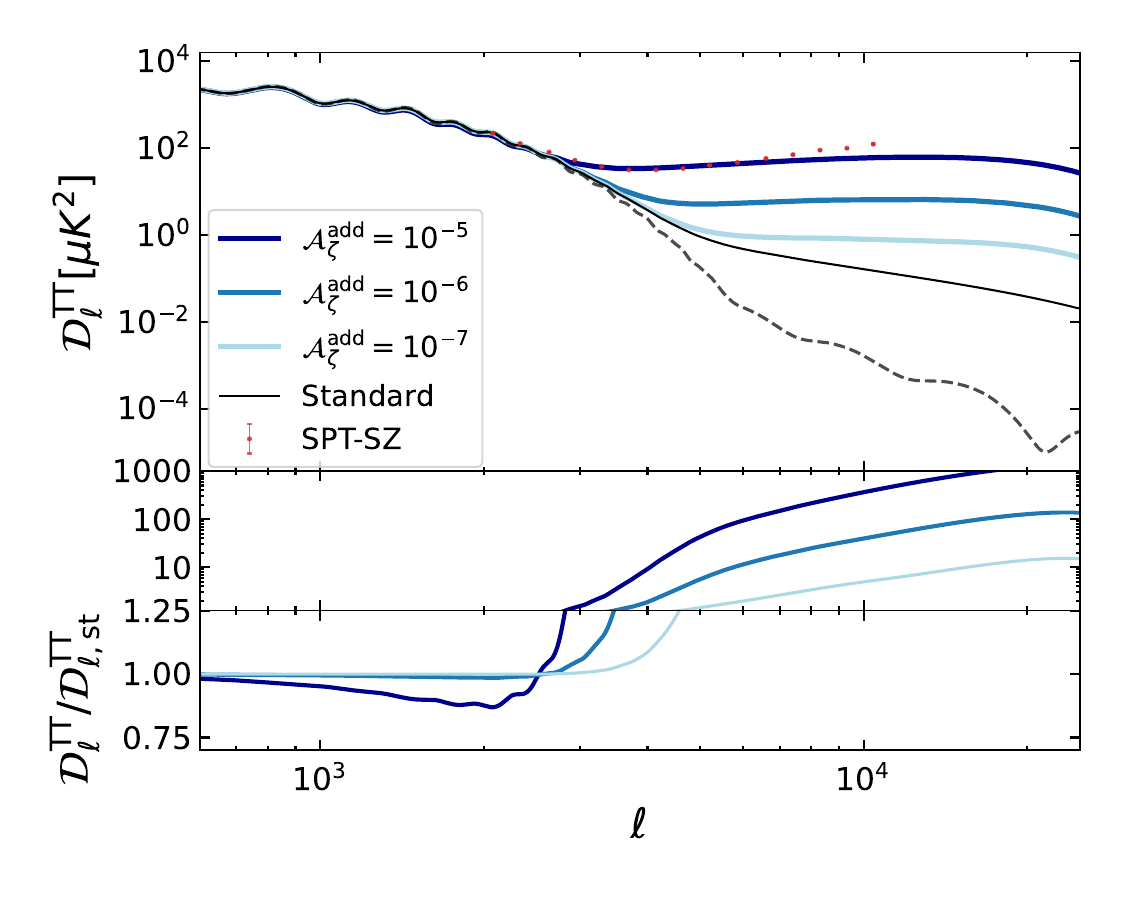}
      \end{minipage} \\
   
      \begin{minipage}[t]{.5\linewidth}
        \centering
        \includegraphics[width=.9\hsize]{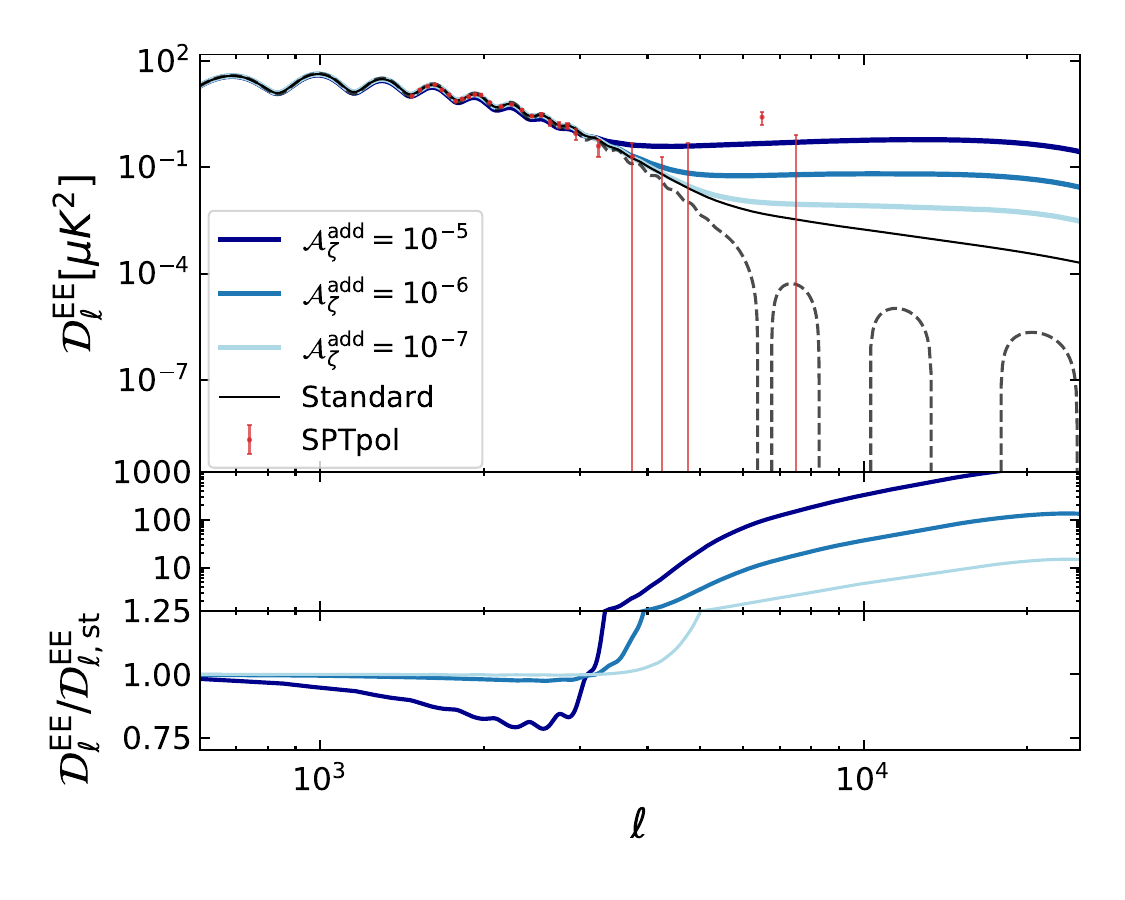}
      \end{minipage} &
      \begin{minipage}[t]{.5\linewidth}
        \centering
        \includegraphics[width=.9\hsize]{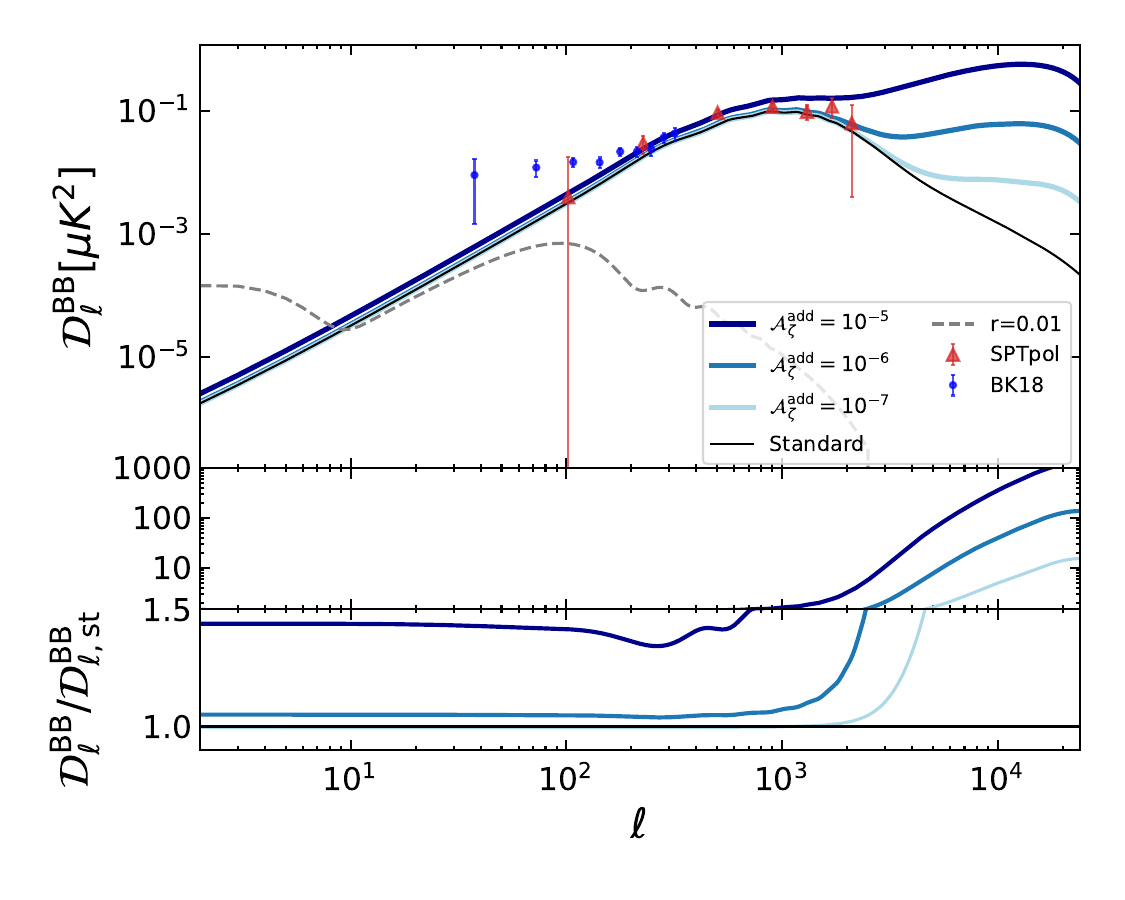}
      \end{minipage} 
    \end{tabular}
    \caption{The angular power spectrum of the lensing potential and the lensed angular power spectrum of the CMB temperature, E-mode, and B-mode polarization with three models, $\mathcal{A}_{\zeta}^{\mr{add}}=10^{-5}$~(dark blue), $10^{-6}$~(blue), and $10^{-7}$~(light blue). We fix the value of $k_{\mathrm{sp}}$, $k_{\mathrm{sp}}=2.6\mathrm{Mpc}^{-1}$. We plot these spectrums in the same manner as Figs.~\ref{fig: clpp_comparison} and \ref{fig: cls_M1e10}. \textit{Upper left}: The angular power spectrum of the lensing potential,
    \textit{Upper right}: The lensed angular power spectrum of the CMB temperature, 
    \textit{Lower left}: The lensed angular power spectrum of the CMB E-mode polarization, 
    \textit{Lower right}: The lensed angular power spectrum of the CMB B-mode polarization.}
    \label{fig: cls_Amat_dependencies}
  \end{figure*}

We show dependencies of $C_{\ell}^{\phi\phi}$ and the lensed CMB angular power spectrum such as $C_\ell^{\mathrm{L},TT}$, $C_\ell^{\mathrm{L},EE}$, and $C_\ell^{\mathrm{L},BB}$ on $\mathcal{A}_{\zeta}^{\mr{add}}$. 
We fix here the value of $k_{\mathrm{sp}}$ as $k_{\mathrm{sp}}=2.6\mr{Mpc}^{-1}$ and calculate for three models, $\mathcal{A}_{\zeta}^{\mr{add}}=10^{-5}$, $10^{-6}$, and $10^{-7}$. 
We presented an $\mathcal{A}\zeta^{\mathrm{add}}=10^{-5}$ model as one example; however, note that the amplitude of $\mathcal{A}\zeta^{\mathrm{add}}=10^{-5}$ has already been ruled out from the COBE/FIRAS observations.
Figure~\ref{fig: cls_Amat_dependencies} shows the dependencies plotted in the same manner as in Figs.~\ref{fig: clpp_comparison} and \ref{fig: cls_M1e10}.
The upper left panel shows $C_{\ell}^{\phi\phi}$ represented by the dark blue line for the model of $\mathcal{A}_{\zeta}^{\mr{add}}=10^{-5}$, the blue line for $\mathcal{A}_{\zeta}^{\mr{add}}=10^{-6}$, and the light blue for $\mathcal{A}_{\zeta}^{\mr{add}}=10^{-7}$. The black solid and dotted lines show the standard angular power spectrum with and without lensing effects. 
We can find that the model of $\mathcal{A}_{\zeta}^{\mr{add}}=10^{-5}$ would also be ruled out through the measurement of $C_{\ell}^{\phi\phi}$.
At the top of the other three panels in Fig.~\ref{fig: cls_Amat_dependencies}, we plot the lensed CMB angular power spectrum of $C_\ell^{\mathrm{L}, TT}$~(upper right), $C_\ell^{\mathrm{L}, EE}$~(lower left), and $C_\ell^{\mathrm{L}, BB}$~(lower right), including the nonlinear effect with EFHs and standard DM halos as in Fig.~\ref{fig: cls_M1e12}.
At the bottom of these three panels, we show the ratios of the solid lines against the black solid line as in Fig.~\ref{fig: cls_M1e10}.
We find at most $\sim 50\%$ increments of the amplitude of the CMB $B$-mode polarization anisotropy with the model of $\mathcal{A}_{\zeta}^{\mr{add}}=10^{-5}$ on large scales, while the enhancement with the model of $\mathcal{A}_{\zeta}^{\mr{add}}=10^{-7}$ is too small to detect.

\section{$z_{\mr{suv}}$ dependency}\label{appendix: zsuv}
\begin{figure*}[htbp]
    \begin{tabular}{cc}
      \begin{minipage}[t]{.5\linewidth}
        \centering
        \includegraphics[width=.9\hsize]{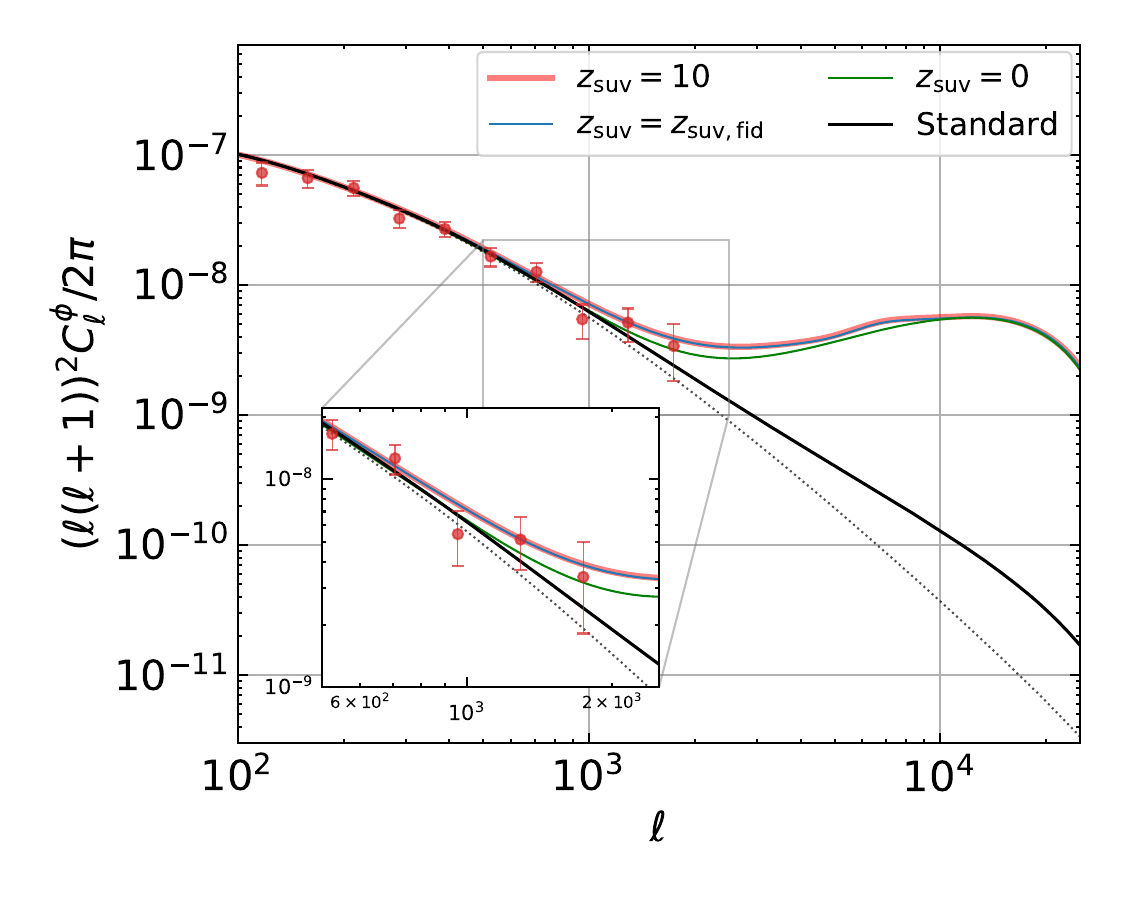}
      \end{minipage} &
      \begin{minipage}[t]{.5\linewidth}
        \centering
        \includegraphics[width=.9\hsize]{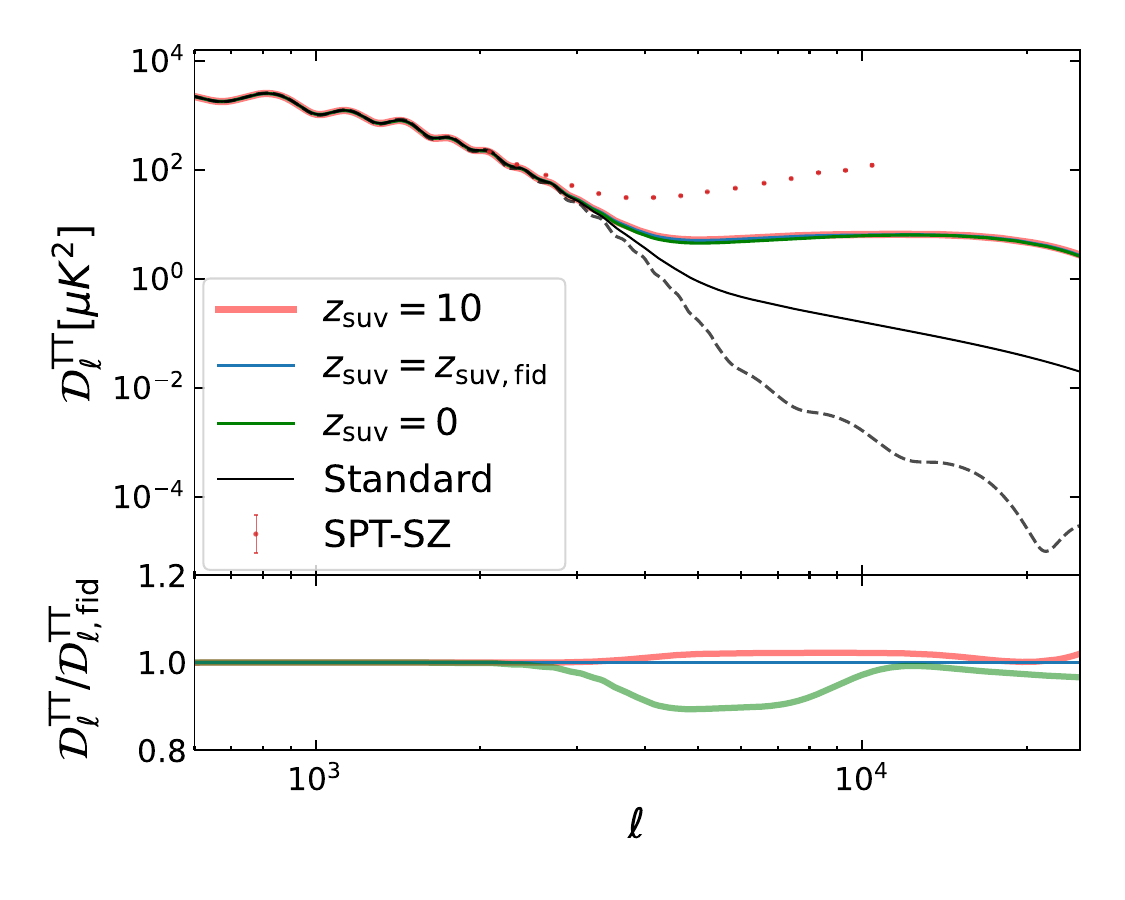}
      \end{minipage} \\
   
      \begin{minipage}[t]{.5\linewidth}
        \centering
        \includegraphics[width=.9\hsize]{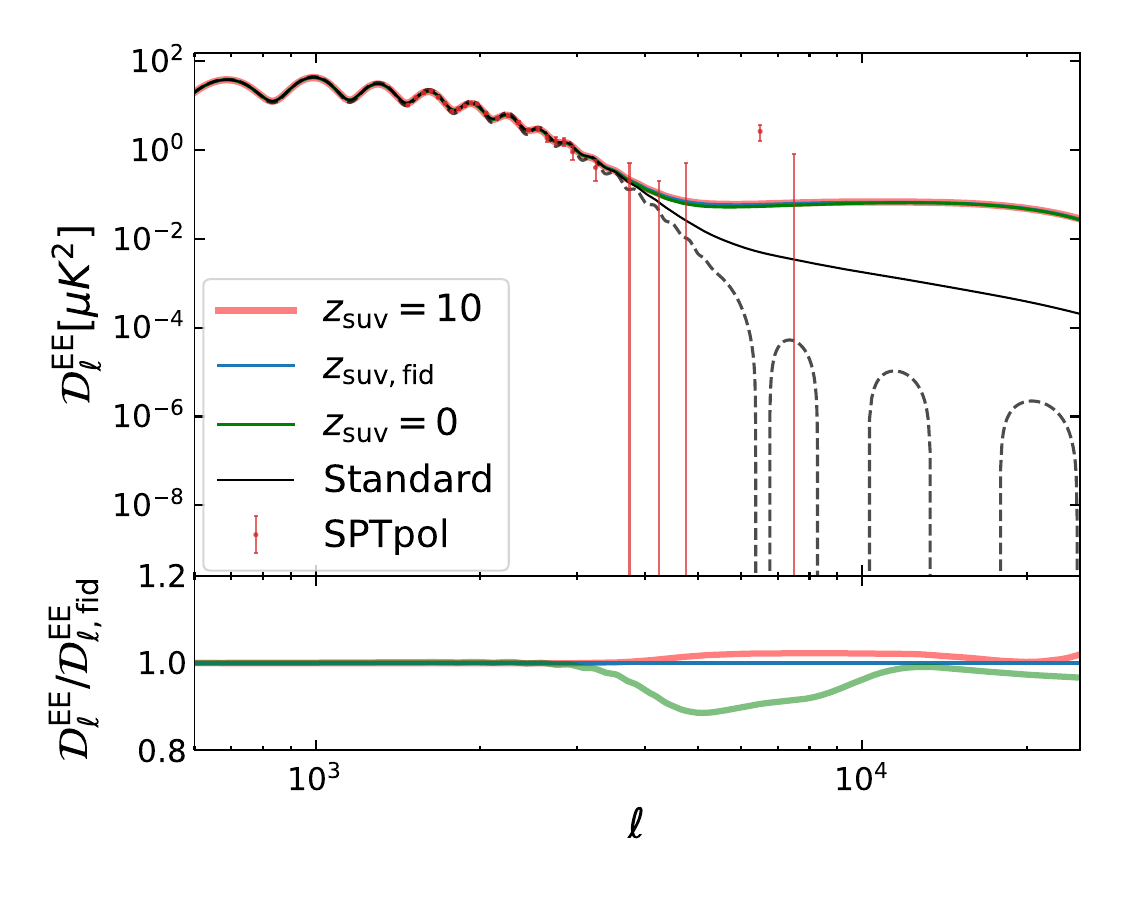}
      \end{minipage} &
      \begin{minipage}[t]{.5\linewidth}
        \centering
        \includegraphics[width=.9\hsize]{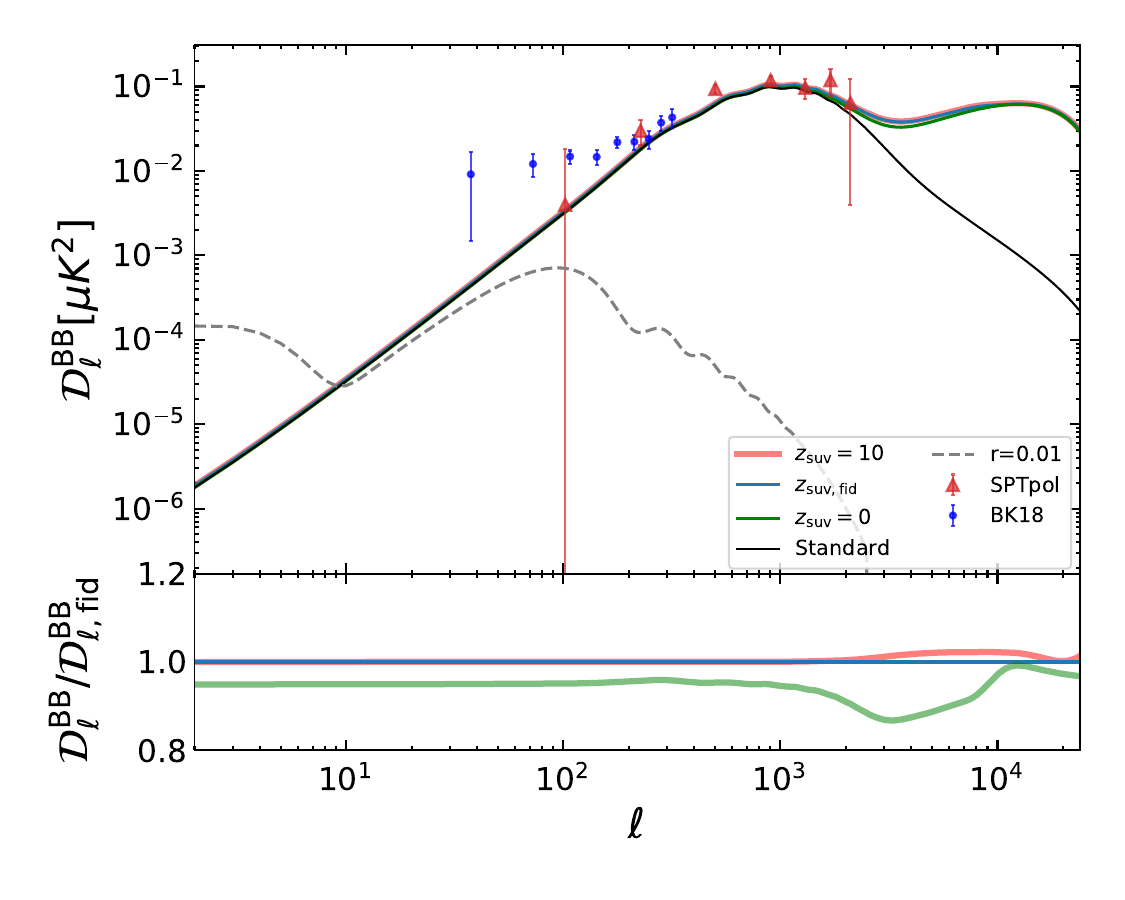}
      \end{minipage} 
    \end{tabular}
    \caption{The angular power spectrum of the lensing potential and the lensed angular power spectrum of the CMB temperature, E-mode, and B-mode polarization of the model of $(k_{\mathrm{sp}}~[\mathrm{Mpc}^{-1}], \mathcal{A}_{\zeta}^{\mathrm{add}})=(2.6, 10^{-6})$ with three different $z_{\mathrm{suv}}$ of $z_{\mathrm{fid}}\approx 3$~(blue), $10$~(red wide) and $0$~(green). We plot these spectrums in the same manner as Figs.~\ref{fig: clpp_comparison} and \ref{fig: cls_M1e12}. \textit{Upper left}: The angular power spectrum of the lensing potential,
    \textit{Upper right}: The lensed angular power spectrum of the CMB temperature, 
    \textit{Lower left}: The lensed angular power spectrum of the CMB E-mode polarization, 
    \textit{Lower right}: The lensed angular power spectrum of the CMB B-mode polarization.}
    \label{fig: cls_zsuv_dependencies}
  \end{figure*}

We show dependencies of of $C_{\ell}^{\phi\phi}$ and the lensed CMB angular power spectrum such as $C_\ell^{\mathrm{L},TT}$, $C_\ell^{\mathrm{L},EE}$, and $C_\ell^{\mathrm{L},BB}$ on $z_{\mr{suv}}$.
Here, we fix the values of $k_{\mathrm{sp}}$ and $\mathcal{A}_{\zeta}^{\mr{add}}$ as $(k_{\mr{sp}}[\mr{Mpc}^{-1}],\mathcal{A}_{\zeta}^{\mathrm{add}})=(2.6,10^{-6})$.
In addition to our fiducial value, \textit{i.e.,} $z_{\mr{suv}}^{\mr{fid}}\approx 3$ as described in the main text, calculations are performed for $z_{\mr{suv}} = 10$ and $z_{\mr{suv}} = 0$.
Figure~\ref{fig: cls_zsuv_dependencies} shows the dependencies.
The upper left panel shows $C_{\ell}^{\phi\phi}$ represented by the blue line for the fiducial calculation, by the red line for the case of $z_{\mr{suv}} = 10$, and by the green line for the case of $z_{\mr{suv}} = 0$. The black solid and dotted lines show the standard angular power spectrum with and without lensing effects. 
At the top of the other three panels in Fig.~\ref{fig: cls_zsuv_dependencies}, we plot the lensed CMB angular power spectrum of $C_\ell^{\mathrm{L}, TT}$~(upper right), $C_\ell^{\mathrm{L}, EE}$~(lower left), and $C_\ell^{\mathrm{L}, BB}$~(lower right), including the nonlinear effect with EFHs and standard DM halos as in Fig.~\ref{fig: cls_M1e12}.
At the bottom of these three panels, we show the ratios of the solid lines against the blue solid line, \textit{i.e.,} the fiducial spectrum.
We find that an increment of the value of $z_\mathrm{suv}$ slightly increases the signals for the model. We also find that reducing $z_\mathrm{suv}$ weakens the signals. Although we do not show explicitly, we performed the same calculations for the other two models studied in the main text and confirmed that an increment in $z_\mathrm{suv}$ does not affect CMB observables.
This is because the negative effect of not considering the nonlinear effects due to the standard halos is larger than the addition of the EFHs' nonlinear effects at lower redshifts. This indicates that the nonlinear effects of the standard halos are more dominant than those of the EFHs.


\bibliography{article}
\end{document}